\documentclass[graybox]{svmult}

\usepackage{mathptmx}       % selects Times Roman as basic font
\usepackage{helvet}         % selects Helvetica as sans-serif font
\usepackage{courier}        % selects Courier as typewriter font
\usepackage{type1cm}        % activate if the above 3 fonts are
\usepackage{makeidx}         % allows index generation
\usepackage{graphicx}        % standard LaTeX graphics tool
\usepackage{multicol}        % used for the two-column index
\usepackage[bottom]{footmisc}% places footnotes at page bottom

\usepackage{amsmath,amsthm,amsfonts,amssymb,alltt}

\makeindex             % used for the subject index
\begin{document}

\title*{Effective Conductivity and Critical Properties of a Hexagonal Array of Superconducting Cylinders}
\author{Simon Gluzman, Vladimir Mityushev, Wojciech Nawalaniec, Galina Starushenko}
\institute{Simon Gluzman \at Ekayna Vihara,                                  Bathurst 3000, Apt 606, Toronto ON, M6B 3B4, Canada \& Ryukoku University, Kyoto 
 \\ \email{simon.gluzman@gmail.com}
\and Vladimir V. Mityushev \at Dept. Computer Sciences and Computer Methods, Pedagogical University, ul. Podchorazych 2, Krakow 30-084, Poland  \\ \email{mityu@up.krakow.pl}
\and Wojciech Nawalaniec \at Dept. Computer Sciences and Computer Methods, Pedagogical University, ul. Podchorazych 2, Krakow 30-084, Poland  \\ \email{wnawalaniec@gmail.com}
\and Galina Starushenko \at Dnipropetrovs'k Regional Institute of State Management of National Academy of State Management at the President of Ukraine Gogolya 29, UA-49631, Dnipropetrovs'k, Ukraine   \\ \email{gs\_gala-star@mail.ru}}

% Use the packa ge "url.sty" to avoid
\maketitle

\abstract{
Effective conductivity of a 2D composite corresponding to the regular hexagonal arrangement of superconducting disks is expressed in the form of a long series in the volume fraction of ideally conducting
disks. According to our calculations based on various re-summation techniques, both the threshold and critical index are obtained in good agreement with expected values. The critical amplitude is in the interval 
$(5.14,5.24)$ that is close to the theoretical estimation $5.18$. The next order (constant) term in the high concentration regime is calculated for the first time, and the best estimate is equal to $(-6.229)$. Final formula is derived for the effective conductivity for arbitrary volume fraction.}

\section{Introduction}
We consider a two-dimensional composite corresponding to the regular hexagonal lattice arrangement of  ideally conducting (superconducting) cylinders of radius $a$ embedded into the matrix of  a conducting material.
The studies of the effective conductivity $\sigma(x)$ of regular composites were pioneered  by Maxwell \cite{0} and Rayleigh \cite{1}. The results of these fundamental research remained limited  to the lowest orders in $x$. Their work was extended in \cite{McPh}, resulting in rather good numerical solutions valid in much broader concentration intervals. 

The effective conductivity $\sigma(x)$ is an analytic function in $x$. In general case of a two-phase composite the so-called contrast parameter should be also included into consideration explicitly, see e.g. \cite{m3}. 
We are interested in the case of a high contrast regular composites, when the conductivity of the inclusions is much larger than the conductivity of the host.  I.e., the highly conducting inclusions are replaced by the ideally conducting inclusions with infinite conductivity. In this case the contrast parameter is equal to unity and remains implicit. The conductivity of the matrix is normalized by unity as well. From the phase interchange theorem \cite{kell0} it follows that in two-dimensions  the superconductivity problem is dual to the conductivity problem, and the superconductivity critical index is equal to the conductivity index.

Our study is restricted  to the two-dimensional case which is still interesting, both for practical \cite{LB,hexa1} and physical reasons \cite{McPh,torq}. Composite materials often consist of a uniform background-host reinforced by a large number (high concentration) of unidirectional rod -or  fiber-like inclusions with high conductivity \cite{hexa1}. 

On the other hand, two-dimensional regular hexagonal- arrayed  composites \cite{hexa1}, much closer resemble the two-dimensional random composites, than their respective 3D counterparts do \cite{torq}. The tendency to order in the two-dimensional random system of disks, is a crucial feature in the theory of composites at high concentrations. 

Most strikingly  it appears that  the maximum volume fraction of  $\frac{\pi}{\sqrt{12}}\approx 0.9069$, 
is attained both for the regular hexagonal array of disks and for random (irregular) 2D composites \cite{torq}. 

A numerical study of the 2D hexagonal case can be found in \cite{McPh}. Their final formula (\ref{proxy}) 
\begin{equation}
\label{proxy}
\sigma(x)=1-\frac{2 x}{\frac{0.075422 x^6}{1-1.06028 x^{12}}+x-1} ,
\end{equation}
compares rather well with numerical data of \cite{McPh}. Note that (\ref{proxy}) diverges with critical exponent $ s=1$, as $x\rightarrow0.922351$. This property on one hand makes the formula more accurate in the vicinity of a true critical point but, on the other hand, makes any comparison in the critical region meaningless. It remains rather accurate till $x=0.85$, where the error is $0.47\%$. For $x=0.905$ the error is $52\%$.
 Expression (\ref{proxy}) was derived using only terms up to the 12th order in concentration. The expansion of (\ref{proxy}) is characterized by a rather regular behavior of the coefficients,

\begin{equation}
\label{seriesreg1}
\begin{array}{llll}
\sigma^{reg}{(x)}=
1+2 x+2 x^2+2 x^3+2 x^4+2 x^5+2 x^6+
\\
2.15084 x^7+2.30169 x^8+2.45253 x^9+2.60338 x^{10}+
\\
2.75422 x^{11}+2.90506 x^{12}+O(x^{13}).
%\\3.06728 x^{13}+3.24088 x^{14}+
%\\
%%3.42586 x^{15}+3.62221 x^{16}+3.82994 x^{17}+4.04904 x^{18}+
%%\\
%%4.44032 x^{19}+4.84469 x^{20}+5.26302 x^{21}+5.69615 x^{22}+
%%\\
%%6.14495 x^{23}+6.61027 x^{24}+7.11717 x^{25}+7.66663 x^{26}+
%%\\
%%O(x^{27}).
\end{array}
\end{equation}

%VM
One can, in principle, collect the higher-order terms as well. However, such derivation of an additional terms can not be considered as consistent since it relies on the agreement with numerical results.
It turns out though, that (\ref{seriesreg1}) compares well with our results shown below, see (\ref{series}).
Except an immediate vicinity of the critical point, analytic-numeric approach of \cite{hexa1}, is in a good agreement with the numerical results of \cite{McPh}.

%VM
In a different limit of high concentrations Keller \cite{kell1}  suggested a constructive asymptotic method for regular lattices, leading to very transparent, inverse square-root formula for the square array \cite{kell1}. Berlyand and Novikov \cite{BN} extended Keller's method to the hexagonal array,
\begin{equation}
\sigma\simeq \frac{\sqrt[4]{3} \pi^{3/2}}{\sqrt{2}} \frac{1}{\sqrt{\frac{\pi}{\sqrt{12}}-x}}. \label{kellcrit}
\end{equation}
%which matches the general critical form and extends the result obtained by Keller for the square lattice \cite{kell1}, by means of a simple multiplication by $\sqrt{3}$, 
Thus the  critical amplitude $A$ (pre-factor), is equal to $A \approx 5.18$.

We will examine below this result for the critical amplitude from the perspective of re-summation techniques suggested before for square regular arrays \cite{our1}. By analogy with square lattice \cite{MCass1}, we expect a constant correction in the asymptotic regime,
\begin{equation}
\sigma\simeq \frac{\sqrt[4]{3} \pi^{3/2}}{\sqrt{2}} \frac{1}{\sqrt{\frac{\pi}{\sqrt{12}}-x}}+B, \label{kellcrit2}
\end{equation}
where the correction term $B$ can not be found in the literature, to the best of our knowledge. It will be calculated below by different methods.

With account for such correction, the final universal formula valid for all possible concentrations from $0$ to $\frac{\pi}{\sqrt{12}}$, has the form
\begin{equation}
\label{final}
\sigma{(x)}=a(x)\; \frac{P(x)}{Q(x)},
\end{equation}
where 
$$
a(x)= \frac{36.1415}{\sqrt{\frac{\pi}{\sqrt{12}}-x}} + 15.9909\sqrt{\frac{\pi}{\sqrt{12}}-x}-45.685+  2.46148 x,
$$
$$
P(x)= (0.939152 + x)(1.38894 - 2.16685 x + x^2)\times 
$$
$$
 (2.55367 - 0.836613 x + x^2)(2.08347 + 2.12786 x + x^2)
$$
and
$$
Q(x)= (1.01215 + x) (1.61369 - 2.31669 x + x^2) \times 
$$
$$
(6.51762 - 0.173965 x + x^2) (4.88614 + 3.28716 x + x^2).
$$

%We will examine below the O'Brien conjecture for the critical amplitude from the perspective of re-summation techniques suggested for the square lattice regular  arrays \cite{our1}. It turns out that it should be corrected by an appropriate geometrical factor.
The rest of the paper is organized as follows: 
in Section~\ref{seriesreg} we describe the essentials of the long series derivation.
Section~\ref{critpoint}, applies various methods to the critical point calculation and compares
the obtained results. 
In Section~\ref{indamp} the critical index and amplitude $A$ are calculated.
Section~\ref{ampformul} where the most accurate formula for all volume fractions is derived, comparing the obtained predictions to numerical data. The amplitude $B$ is calculated.
Section~\ref{high} is concerned with interpolation with Pade approximants.
Section~\ref{ansatz} returns to discussion of the ansatz for construction of the starting approximation.
Section~\ref{united} gives unified approach to the square and hexagonal lattices.
Section~\ref{Largen} considers Dirichlet summation to extract the asymptotic behavior of series coefficients.
Section~\ref{lubrication} derives the asymptotic formula by use of the lubrication theory.
Section~\ref{RAND} considers random composites related to the hexagonal lattice.
Finally, Section~\ref{discussion} concludes with a discussion of obtained results. 

\section{Series for Hexagonal array of superconducting cylinders
\label{seriesreg}
}
We proceed to the case of a hexagonal lattice of inclusions, where rather long expansions in concentration  will be presented an analyzed systematically.
 The  coefficients $a_{n}$ in the expansion of $\sigma(x)=1+\sum _{n=1}^{\infty} a_n x^n$, are expressed through elliptic functions by exact formulas from \cite{m1,m2}. Below, this expansion is presented in the truncated numerical form,
\begin{eqnarray}
\label{series}
\sigma{(x)}=&&
1 + 2 x + 2 x^2 + 2 x^3 + 2 x^4 +2 x^5 +2 x^6
\nonumber\\
&+&
2.1508443464271876 x^7 +2.301688692854377 x^8
\nonumber\\
 &+& 
2.452533039281566 x^9 +2.6033773857087543 x^{10}
 \nonumber\\
 &+&
2.754221732135944 x^{11} +2.9050660785631326 x^{12} 
 \nonumber\\
 &+&
3.0674404324522926 x^{13}  +3.2411917947659736` x^{14} 
 \nonumber\\
&+&
3.426320165504177 x^{15} +3.6228255446669055 x^{16}
\nonumber\\
&+& 
3.8307079322541555 x^{17} +4.049967328265928 x^{18} 
\nonumber\\ 
 &+&
 4.441422739726373 x^{19} +4.845994396051242 x^{20} 
 \nonumber\\ 
 &+&
5.264540375940583 x^{21} +5.69791875809444 x^{22} 
 \nonumber\\ 
&+& 
6.146987621212864 x^{23} +6.6126050439959 x^{24}
\nonumber\\ 
 &+&
7.135044602470776 x^{25}+7.700073986554016 x^{26} 
 \nonumber\\ 
&+& 
O(x^{27}).
\end{eqnarray}
The first twelve coefficients of \eqref{series} and the Taylor expansions of (\ref{proxy}) coincide. The next coefficients can be calculated by exact formulas from \cite{m1,m2}. This requires use of the double precision and perhaps a power computer, not a standard laptop.  

Since we are dealing with the limiting case of a perfectly conducting inclusions when the conductivity of inclusions tends to infinity, the effective conductivity is also expected to tend to infinity as a power-law, as the concentration $x$ tends
to the maximal value $x_{c}$ for the hexagonal array,
\begin{equation}
\label{crit}
\sigma(x)\simeq A  (x_c-x)^{-s} +B.
\end{equation}
The critical superconductivity index (exponent) $s$ is believed to be  $1/2$ for all lattices \cite{BN}. For sake of exploring how  consistent are various resummation techniques, we will calculate the index.  The critical amplitudes $A$ and $B$ are unknown non-universal parameters to be calculated below as well.

The problem of interest can be formulated mathematically as follows. Given the polynomial approximation \eqref{series} of the function $\sigma(x)$, to estimate the convergence radius $x_c$ of the Taylor series $\sigma(x)$; to determine critical index $s$ and amplitudes $A, B$ of the asymptotically equivalent approximation (\ref{crit}) near $x=x_c$.

 When such extrapolation problem is solved, we proceed to solve an interpolation problem of matching the two asymptotic expressions for the conductivity and  derive interpolation formula for all concentrations.

\section{Critical Point
\label{critpoint}
}
\subsection{Pad\'{e} approximants}

Probably the simplest and direct way to extrapolate, is to apply the Pad\'{e} approximants $P_{n,m}(x)$, which is nothing else but ratio of the two polynomials $P_{n}(x)$ and $P_{m}(x)$ of the order $n$ and $m$, respectively. The coefficients are derived directly from the coefficients of the given power series \cite{pade1, Suetin} from the requirement of asymptotic equivalence to the given series or function $f(x)$. When there is a need to stress the last point, we simply write $PadeApproximant[f[x],n,m]$. 

In order to estimate the position of a critical point, let us apply the diagonal  Pad\'{e} approximants,
\begin{equation}
P_{1,1}(x)=\frac{m_1 x+1}{n_1 x+1}, \quad P_{2,2}(x)=\frac{m_2 x^2+m_1 x+1}{n_2 x^2+n_1 x+1}, \ldots
\end{equation}
Pad\'{e} approximants locally are the best rational approximations of power series. Their poles determine singular points of the approximated functions \cite{Suetin,pade1}. Calculations with Pad\'{e} approximants are straightforward and can be performed with $Mathematica^{\textsuperscript{\textregistered}}$. 
They do not require any preliminary knowledge of the critical index and we have to find the position of a simple pole.  In the theory of periodic 2D composites \cite{moment,moment1,moment2}, their application is justifiable rigorously  away from the square-root singularity and from the high-contrast limit.

There is a convergence within the approximations for the critical point generated  by the sequence of  Pad\'{e} approximants, corresponding to their order increasing:
\\
$x_1=1$, $x_2=1$, $x_3=1$, $x_4-n.a,$, $x_5 -n.a.$,
$x_6=0.945958$, $x_7=0.945929$, $x_8= 0.947703$, $x_9= 0.946772$, $x_{10}=0.942378$, 
$x_{11}=0.945929$, $x_{12}= 0.945959$, $x_{13}=0.920878$.

The main body of the approximations is well off the exact value. The percentage error given by the last/best approximant in the sequence equals to  $1.5413\%$. If only the first row of the  Pad\'{e} table is studied \cite{Suetin}, then the best estimate is equal to $0.929867$, close to the estimates with the diagonal sequence.

We suggest that further increase in accuracy is limited by triviality, or 
``flatness'' of the coefficients values in six starting orders of (\ref{series}).  Consider another sequence of approximants, when diagonal Pad\'{e} approximants are multiplied with Clausius-Mossotti-type expression, 
\begin{eqnarray}
P_{1}^{t} (x)=\frac{(1-x)}{(1+x)} \frac{(1+m_1 x)}{(1+n_1 x)};
\nonumber\\
P_{2}^{t} (x)=\frac{(1-x)}{(1+x)} \frac{(1+m_1 x+m_2 x^2)}{(1+n_1 x+n_2 x^2)},...
\end{eqnarray}
The transformation which lifts the flatness, does improve convergence of the sequence of approximations for the threshold,
\\
$x_7=0.94568$, $x_8-n.a.$, $x_9=0.948299$,  $x_{10}=0.9287$, $x_{11}=0.945681$, $x_{12}=0.89793$, $x_{13}=0.903517$.
The percentage error given by the last approximant in the sequence equals  $-0.373\%$.

In order to judge the quality of the latter estimate, let us try highly recommended $D-Log$ Pad\'{e} method \cite{pade1}, which also does not require a preliminary knowledge of the critical index value. One has to differentiate $Log$ of (\ref{series}), apply the diagonal  Pad\'{e} approximants and define the critical point as the position of the pole nearest to the origin. The best estimate obtained this way is $x_{12}=0.919304$, with percentage error of $1.368\%$. One can also estimate the value of critical index as a residue \cite{pade1}, and obtain rather disappointing value of $0.73355$.

\subsection{Corrected Threshold}

An approach based on the Pad\'{e} approximants produces  the expressions for the cross-properties from "left-to-right", extending the series from the dilute regime of small $x$ to the high-concentration regime of large $x$. Alternatively, one can proceed from "right-to-left", i.e. extending the series from the large $x$ (close to $x_{c}$) to small $x$ \cite{our1,cross,cross1}. 

 We will first derive an approximation to the high-concentration regime and then extrapolate to the less concentrated regime. There is an understanding  that physics of a 2D high-concentration, regular and irregular composites is related to the so-called "necks", certain areas between closely spaced disks \cite{kell1,LB,BN}. 

Assume also that the initial guess for the threshold value is available from previous Pad\'{e} -estimates, and is equal to $x_6=0.945958$.

The simplest way to proceed is to look for the solution in the whole region $[0,x_{c})$,
in the form which extends asymptotic expression from \cite{MCass}, 
$\sigma=\alpha_{1} (x_{c}-x)^{-1/2}+\alpha_{2}$. This approximation works well for the square lattice of inclusions \cite{our1}.

 In the case of hexagonal lattice we consider its further extension,

\begin{eqnarray}
\label{and}
\sigma=\alpha_{1} (x_{6}-x)^{-s}+\alpha_{2} +\alpha_{3} (x_{6}-x)^{s},
\end{eqnarray}
where index s is considered as another unknown. All unknowns can be obtained from the three starting non-trivial terms of (\ref{series}), namely $\sigma\simeq 1+2x+2 x^2+2x^3$. Thus the parameters equal $\alpha_{1}=2.24674$, $\alpha_{2}=-1.43401$, $\alpha_{3}=0.0847261$, $s=0.832629$.

Let us assume that the true solution $\sigma$ may be found in the same form but with exact, yet unknown threshold $X_{c}$,

\begin{equation}
\label{exact}
\Sigma=\alpha_{1} (X_{c}-x)^{-s}+\alpha_{2} +\alpha_{3} (X_{c}-x)^{s}.
\end{equation}
The expression (\ref{exact}) may be inverted and $X_{c}$ expressed explicitly,
\begin{equation}
\label{th}
X_c=2^{-1/s} \left(\frac{-\sqrt{\left(\alpha _2-\Sigma \right){}^2-4 \alpha _1 \alpha _3}-\alpha _2+\Sigma }{\alpha _3}\right){}^{1/s}+x.
\end{equation}

Formula (\ref{th}) is a formal expression for the threshold, since $\Sigma(x)$ is also unknown. We can use for $\Sigma$ the series in $x$, so that instead of a true threshold we have an effective threshold, $X_{c}(x)$, given in the form of a series in $x$. For the concrete series (\ref{series}), the following expansion follows,

\begin{eqnarray}
\label{series0}
X_{c}(x)=&& x_6 +0.0134664 x^4+0.00883052 x^5 
\nonumber\\
&+&
0.00647801 x^6-0.0709217 x^7 +0.0032732 x^8 
\nonumber\\
&+&
0.00244442 x^9+0.00594779 x^{10}+0.00482187 x^{11}
\nonumber\\
&+&
0.00413887 x^{12}+...,
\end{eqnarray}
which should become a true threshold $X_{c}$ as $x\rightarrow X_{c}$.

Moreover, let us apply re-summation procedure to the expansion 
(\ref{series0}) using the diagonal Pad\'{e} approximants. Finally let us define the sought threshold $X_{c}^{*}$ self - consistently from the following equations dependent on the approximants order,

\begin{equation}
\label{thpa0}
X_{c,n}^{*}=0.945958+0.0134664 x^4 P_{n,n}(X_{c}^{*}),
%PadeApproximant[X_{c}(x),n,m].
\end{equation}
meaning simply that as we approach the threshold, the RHS of (\ref{thpa0}) should become the threshold. Since the diagonal Pad\'{e}  approximants of the $n$-th order are defined for an even number of terms $2n$, we will also have a sequence of $X_{c,n}^{*}$. 

Solving equation (\ref{thpa0}), we obtain $X_{c,4}^{*}=0.930222$,  $X_{c,5}^{*}=0.855009$,  $X_{c,6}^{*}=0.9483$, $X_{c,7}^{*}=0.932421$, $X_{c,8}^{*}=0.946773$, $X_{c,9}^{*}=0.941391$. $X_{c,10}^{*}=0.94682$, $X_{c,11}^{*}=0.932752$, $X_{c,12}^{*}=0.907423$, $X_{c,13}^{*}=0.903303$. The last two estimates for the threshold  are good.

\subsection{Critical index is known}

Also, one can pursue a slightly different strategy, assuming that critical index is known ($s=1/2$), and is incorporated into initial approximation. Recalculated  parameters equal $\alpha_{1}=5.12249$, $\alpha_{2}=-5.74972$, $\alpha_{3}=1.52472$. For the series (\ref{series}), the following expansion follows,

\begin{eqnarray}
\label{series1}
X_{c}(x)=&& x_6 -0.082561  x^3 +0.0282108 x^4-0.000383173 x^5 
\nonumber\\
&+&
0.0228241 x^6-0.0649593 x^7 + 0.01561635 x^8 
\nonumber\\
& -&
0.00911151 x^9+0.01874715 x^{10}+0.00688507 x^{11}
\nonumber\\
&+&
0.0169516 x^{12}+....
\end{eqnarray}

Let us apply re-summation procedure to the expansion 
(\ref{series1})
using super-exponential approximants $E^{*}(x)$ \cite{superexp}. Finally let us define the sought threshold $X_{c}^{*}$ self - consistently, 
\begin{equation}
\label{thr}
X_{c}^{*}=0.945958-0.082561 x^3  E^{*}(X_{c}^{*}).
\end{equation}

Since the super-exponential approximants are defined as $E_{k}^{*}$ for arbitrary number of terms $k$, we will also have a sequence of $X_{c,k}^{*}$.  E.g. 
\begin{eqnarray}
E_{1}^{*}=e^{-0.341697 x},
\nonumber
\\
E_{2}^{*}=e^{-0.341697 e^{0.157266 x} x},
\nonumber
\\
E_{3}^{*}=e^{-0.341697 e^{0.157266 e^{5.28382 x} x} x},...,
\end{eqnarray}
and so on iteratively.  Solving equation ($\ref{thr}$), we obtain $X_{c,1}^{*}=0.901505$, $X_{c,2}^{*}=0.903321$, $X_{c,3}^{*}=0.945958$, $X_{c,4}^{*}=0.903404$,  $X_{c,5}^{*}=0.916641$,  $X_{c,6}^{*}=0.903412$, $X_{c,7}^{*}=0.903556$, $X_{c,8}^{*}=0.903412$, $X_{c,9}^{*}=0.903412$.

There is a convergence in the sequence of approximations for the threshold. The percentage error achieved for the last point is equal to $-0.384537\%$.

The method of corrected threshold produces good results based only on the starting twelve terms from the expansion (\ref{series}), in contrast with the Pad\'{e}-based approximations, requiring all available terms to gain similar accuracy. The task of extracting the threshold, a purely geometrical quantity, from the solution of the physical problem is not trivial and is relevant to similar attempts to find the threshold for random systems from the expressions for some physical quantities \cite{torq}.

Instead of the super-exponential approximants one can exactly as above apply the diagonal  Pad\'{e} approximants, 
\begin{equation}
\label{thrpade}
X_{c,n}^{*}=0.945958-0.082561 x^3  P_{n,n}(X_{c}^{*}).
%PadeApproximant[X_{c}(x),n,m].
\end{equation}
Solving equation ($\ref{thrpade}$), we obtain $X_{c,3}^{*}=0.908188$, $X_{c,4}^{*}=0.889169$,  $X_{c,5}^{*}=0.889391$,  $X_{c,6}^{*}=0.887983$, $X_{c,7}^{*}=0.899495$, $X_{c,11}^{*}=0.903011$, $X_{c,12}^{*}=0.90296$, $X_{c,13}^{*}=0.9057$.

\begin{figure}
\begin{center}
\includegraphics[scale=0.7]{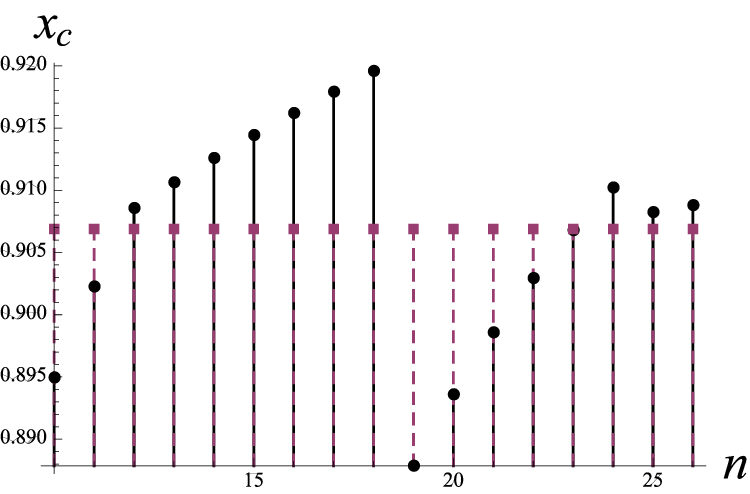}
\end{center}
\caption{
$x_{c}$ calculated by ratio method, compared with the exact threshold.  
}
\label{Figratio}
\end{figure}

Ratio method \cite{pade1}, also works well. It evaluates the threshold through the value of index and ratio of the series coefficients, $x_{c,n}=\frac{\frac{s-1}{n}+1}{\frac{a_n}{a_{n-1}}}$. The last point gives rather good estimate, $x_{c,26}=0.908801$, despite of the oscillations in the dependence on $n$, as seen in Fig.\ref{Figratio}.

\section{Critical Index and Amplitude
\label{indamp}
}
Standard way to proceed with critical index calculations when the value of the threshold is known can be found in  \cite{pade1},\cite{padeour}. One would first apply the following transformation,
\begin{equation}
z=\frac{x}{x_c-x}  \; \Leftrightarrow  \; x=\frac{z x_c}{z+1}, \label{direct}
\end{equation} to the series (\ref{series}) in order to make application of the different approximants more convenient.

Then, to such transformed series $M_{1}(z)$ apply the $D-Log$ transformation 
%(differentiate  $Log$ of $M_{1}(z)$) 
and call the transformed series $M(z)$. In terms of $M(z)$ one can readily obtain the sequence of approximations ${s_{n}}$ for the critical index $s$, 
\begin{equation}
s_{n}=\lim_{z\to \infty } (z PadeApproximant[M[z],n,n+1]).\label{seq1}
\end{equation}
Unfortunately, in the case of (\ref{series}) this approach fails. There is no discernible convergence at all within the sequence of $s_{n}$. Also, even the best result $s_{12}= 0.573035$, is far off the expected $0.5$. Failure of the standard approach underscores the need for a new methods.

\subsection{ Critical Index with $D-Log$ Corrections
\label{shem1}
}
Let us look for a possibility of improving the estimate for the index along the same lines as were already employed in the case of a square lattice of inclusions \cite{our1}, by starting to find a suitable starting approximation for the conductivity and critical index.

Mind that one can derive the  expressions for conductivity  from "left-to-right", i.e. extending the series from small $x$ to large $x$. Alternatively, one can proceed from "right-to-left", i. extending the series from the large $x$ (close to $x_{c}$) to small $x$ \cite{our1,cross,cross1}. 
Let us start with defining reasonable "right-to-left" zero-approximation, which extends the form used in \cite{our1, MCass} for the square arrays. 

The simplest way to proceed is to look for the solution in the whole region $[0,x_{c})$. 
as the formal extension of the expansion,
\begin{equation}
\label{and}
\sigma^{r-l}=\alpha_{1} (x_{c}-x)^{-s}+\alpha_{2}+\alpha_{3}(x_c-x)^{s}+\alpha_{4} (x_c-x)^{2s},
\end{equation}
 All parameters in (\ref{and}) will be obtained by matching it asymptotically with the truncated series $\sigma_{4}=1+2x+2x^2+2x^3+2x^4$, with the following result,
\begin{equation}
\label{general}
\begin{array}{llll}
\sigma_{4}^{r-l}(x)=
\frac{4.69346}{(0.9069\, -x)^{0.520766}}-5.86967+
\\
2.53246 (0.9069\, -x)^{0.520766}-0.526588 (0.9069\, -x)^{1.04153}.
\end{array}
\end{equation}

We present below a concrete scheme for calculating both critical index and amplitude, based on the idea of corrected approximants \cite{corr}. We will attempt to correct the value of  $s_{0}=0.520766$ for the critical index by applying  $D-Log$  Pad\'{e} approximation to the remainder of series (\ref{series}).

 Let us divide the original series (\ref{series}) by $\sigma_{4}^{r-l}(x)$ given by (\ref{general}), apply to the newly found series transformation (\ref{direct}), then apply $D-Log$ transformation and call the transformed series $K(z)$. Finally one can obtain the following sequence of the Pad\'{e} approximations for the corrected critical index,
\begin{equation}
s_{n}=s_{0}+\lim_{z\to \infty } (z PadeApproximant[K[z],n,n+1]) . \label{corindex}
\end{equation}

The following  "corrected" sequence of approximate values for the critical index can be calculated readily:
%$s_1=0.488004$,$s_2=0.478324$,$s_3=0.482625$,
$s_4=0.522573$, $s_5=0.518608$, $s_6=0.554342$, $s_7=0.281015$, $s_8=-0.209639,$, $s_9=0.279669$, $s_{10}=0.527055$, $s_{11}=0.518543$, $s_{12}=0.488502$.The last two estimates surround the correct value.

Generally, one would expect that with adding more terms to the expansion (\ref{series}), quality of estimates for $s$ would improve. As was briefly discussed above, formula (\ref{proxy}) can be expanded in arbitrary order in $x$, generating more terms in expansion (\ref{seriesreg1}).  Such procedure, of course, is not a rigorous derivation of true expansion, but can be used for illustration of the behavior of $s_{n}$ with larger $n$.
%Indeed, in the highest order corresponding to $50$ terms in (\ref{seriesreg1}),  we obtain $s_{25}=0.503267$. Saturation of results is seen in Fig.\ref{Figindex} around this number of terms.

\begin{figure}[t]
\sidecaption[t]
\includegraphics[scale=.65]{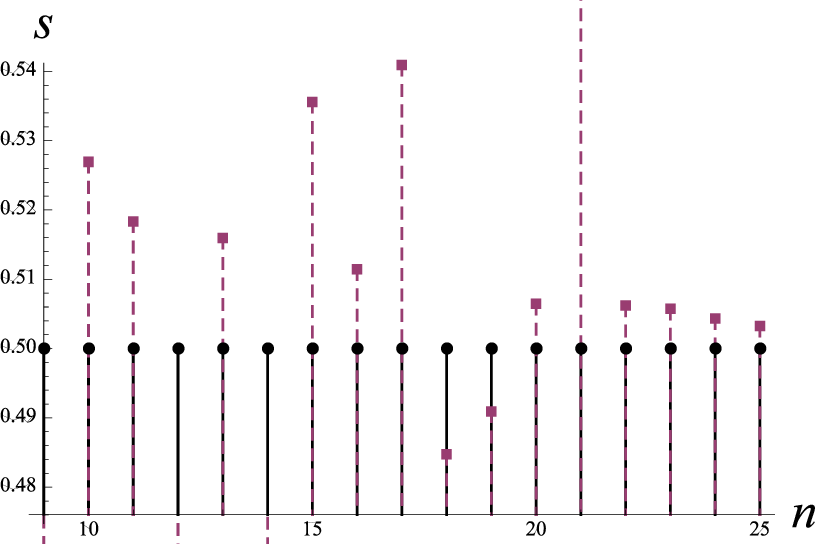}
\caption{Critical index $s$ is calculated by $D-Log$ Corrections method, and compared with the exact value.}
\label{Figindex}     
\end{figure}

If $\gamma_{n}(z)=PadeApproximant[K[z],n,n+1]$, then
\begin{equation}
\label{betaf}
\sigma_{n}^{*}(x)=\sigma_{4}^{r-l}(x)\exp \left(\int_0^{\frac{x}{x_{c}-x}} \gamma_{n}(z)\, dz\right),
\end{equation} 
and one can compute numerically corresponding amplitude,
\begin{equation}
A_{n}=\lim_{x\rightarrow x_{c}}  (x_{c}-x)^{s_{n}}\sigma_{n}^{*}(x),
\end{equation}
with $A_{0}=4.693$. Expressions of the type (\ref{betaf})  have more general form than suggested before in \cite{rg1,puller,padeour}, based on renormalization methods.  

Convergence for the index above is expected to be supplemented by convergence in the sequence of approximate values for critical amplitude, but results are still a bit scattered to conclude about the amplitude value. For the last two approximations we find $A_{11}=4.80599$, $A_{12}=5.38288$, signaling possibility of a larger value than $4.82$, originating from multiplication of the critical amplitude for the square lattice by $\sqrt{3}$, as suggested by O'Brien \cite{McPh}.

To improve the estimates for amplitude $A$, assume that the value of critical index $s=1/2$ is given, and construct $\gamma_{n}(z)$ to satisfy the correct value at infinity. There is now a good convergence for the amplitude, i.e, in the highest orders, $A_{10}=5.09584$, $A_{11}=5.1329$, $A_{12}=5.14063$. Corresponding expression for the approximant
\begin{equation}
\gamma_{12}(z)=\frac{b_{1}(z)}{b_{2}(z)},
\end{equation} 
where
\begin{equation}
\label{b1}
\begin{array}{llll}
b_{1}(z)=
-0.079533 z^4-0.745717 z^5-2.5712 z^6-
\\
4.16091 z^7-2.88816 z^8+0.36028 z^9+
\\
1.74741 z^{10}+0.951728 z^{11}-0.0792987 z^{12},
\end{array}
\end{equation} 
and 
\begin{equation}
\label{b2}
\begin{array}{llll}
b_{2}(z)=1+14.3691 z+94.745 z^2+380.2 z^3+
\\
1037.51 z^4+2036.14 z^5+2961.45 z^6+
\\
3238.1 z^7+2667.9 z^8+1641.88 z^9+
\\
739.461 z^{10}+235.321 z^{11}+48.8016 z^{12}+
\\
3.81868 z^{13}.
\end{array}
\end{equation} 
Corresponding  effective conductivity can be obtained numerically,
\begin{equation}
\label{betaf1}
\sigma_{12}^{*}(x)=\sigma_{4}^{r-l}(x)\exp \left(\int_0^{\frac{x}{x_{c}-x}} \gamma_{12}(z)\, dz\right),
\end{equation} 
The maximum error is at $x=0.905$ and equals $0.4637\%$. It turns out that formula (\ref{betaf1}) is good.

\section{Critical Amplitude and Formula for all Concentrations
\label{ampformul}
}
For practical applications we suggest below the particular re-summation schemes, leading to the analytical expressions for the effective conductivity.
\subsection{Correction with Pad\'{e} approximants
\label{subs}
}
 Let us ensure the correct critical index already in the starting approximation for $\sigma^{r-l}$, so that
all parameters in (\ref{and1}) are obtained by matching it asymptotically with the truncated series $\sigma_{3}=1+2x+2x^2+2x^3$,
\begin{equation}
\label{and1}
\begin{array}{llll}
\sigma_{3}^{r-l}(x)=
\frac{5.09924}{(0.9069\, -x)^{1/2}}-6.67022+
\\
3.04972 (0.9069\, -x)^{1/2}-0.649078 (0.9069\, -x).
\end{array}
\end{equation}

To extract corrections to the critical amplitude, we divide the original series (\ref{series}) by (\ref{and1}), apply to the new series transformation (\ref{direct}). Call the newly found series $G[z]$. Finally  build a sequence of the diagonal Pad\'{e}  approximants, so that the amplitudes are expressed by the formula ($\alpha_{1}=5.09924$),
\begin{equation}
A_{n}=\alpha_{1} \lim_{z\to \infty } ( PadeApproximant[G[z],n,n]) ,
\end{equation} leading to a several reasonable estimates  $A_7=5.26575$, $A_{11}=5.23882$, $A_{12}=5.25781$,
$A_{13}=5.25203$. Complete expression for the effective conductivity corresponding to $A_{11}$ can be reconstructed readily,
\begin{equation}
\label{form1}
\sigma_{11}^{*}(x)=\sigma_{3}^{r-l}(x) C_{11}(x),
\end{equation} 
where $C_{11}(x)=\frac{c_1(x)}{c_2(x)}$,

\begin{equation}
\label{c1}
\begin{array}{llll}
c_{1}(x)=
1.15947+1.13125 x+1.12212 x^2+
\\
1.1167 x^3+3.8727 x^4+0.824247 x^5-
\\
2.62954 x^6+1.19135 x^7+1.21923 x^8+
\\
1.42832 x^9+1.0608 x^{10}+1.53443 x^{11};
\end{array}
\end{equation} 
and
\begin{equation}
\label{c2}
\begin{array}{llll}
c_{2}(x)=
1.15947+1.13125 x+1.12212 x^2+
\\
1.1167 x^3+3.86892 x^4 + 0.849609 x^5 -
\\
2.58112 x^6+1.11709 x^7+1.18377 x^8+
\\
1.36969 x^9+1.06062 x^{10}+x^{11}.
\end{array}
\end{equation} 
Formula (\ref{form1}) is practically as good as (\ref{betaf1}). Maximum error is at the point $x=0.905$ and equals $0.563\%$. 
\subsection{Pad\'{e} approximants. Standard scheme}
Our second suggestion for the conductivity formula valid for all concentrations is based on the following conventional considerations \cite{bender}.
Let us first calculate the critical amplitude $A$. To this end let us again apply transformation (\ref{direct}) to the original series (\ref{series}) to obtain transformed series $M_1(z)$. Then apply to $M_1(z)$ another transformation to get yet another series,
$T(z)=M_{1}(z)^{-1/s}$, in order to get rid of the square-root behavior at infinity. In terms of $T(z)$ one can readily obtain the sequence of approximations ${A_{n}}$ for the critical amplitude $A$, 

\begin{equation}
\label{sug1}
A_{n}=x_{c}^{s}\lim_{z\to \infty } (z PadeApproximant[T[z], {{n, n + 1}}])^{-s};
\end{equation}
There are only few reasonable estimates for the amplitude, $A_6=4.55252$, $A_{11}=4.49882$, $A_{12}=4.64665$ and $A_{13}=4.68505$. The last value is the best if compared with the conjectured in \cite{McPh},  $A=4.82231$.

Following the prescription, the effective conductivity can be easily reconstructed in terms of the Pad\'{e} approximant (corresponding to  $A_{12}$) and compared with the numerical results in the whole region of concentrations. The maximum error is at $x=0.905$, and equals $-5.67482\%$. On the other hand, if the conjectured value $A_{b}$ is enforced  at infinity, through the two-point Pad\'{e} approximant, the results improve and the maximuml error at the same concentration is $-3.18511\%$. Corresponding formula for all concentrations, which also respects 24 terms from the series $T[z]$ is given as follows,

\begin{equation}
\label{form2}
\sigma_{p}^{*}(x)=\frac{1.02555}{\sqrt{0.9069\, -x}} \sqrt{\frac{V_1(x)}{V_2(x)}},
\end{equation}
where

\begin{equation}
\label{v1}
\begin{array}{llll}
V_1(x)=
-0.927562-0.877939 x+0.0406992 x^2+
\\
0.0440014 x^3+0.0414973 x^4+0.0436199 x^5+
\\
0.319848 x^6+0.0110109 x^7-0.122646 x^8+
\\
0.0351069 x^9+0.0439523 x^{10}+0.0380654 x^{11}+
\\
1.01499 x^{12}+x^{13}
\end{array}
\end{equation}

and 
\begin{equation}
\label{v2}
\begin{array}{llll}
V_2(x)=
-1.07571+2.09854 x-2.17187 x^2+
\\
2.23064 x^3-2.3122 x^4+2.374 x^5-
\\
2.1397 x^6+1.87791 x^7-1.78516 x^8+
\\
1.86446 x^9-1.94838 x^{10}+2.03264 x^{11}-
\\
x^{12}
\end{array}
\end{equation}

\begin{figure}
\begin{center}
\includegraphics[scale=0.6]{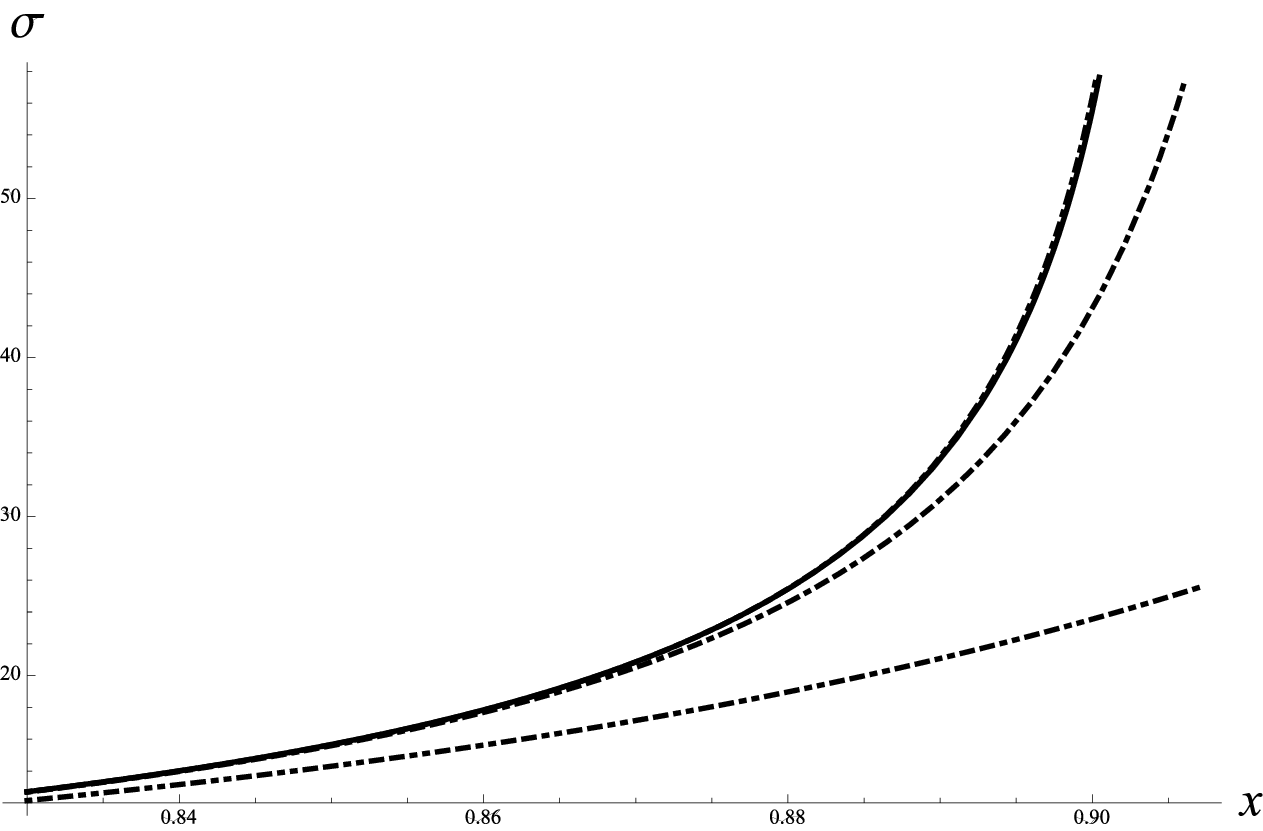}
\end{center}
\caption{
Our formula (\ref{fin4}) (solid) is compared with the 
standard Pad\'{e}  approximant (\ref{form2}) (dotted) and 
rational approximation  (\ref{proxy})  (dashed). The series (\ref{series})  is shown with dashed line.
}
\label{figuren1}
\end{figure}

Various expressions are shown in Fig.\ref{figuren1}. Note, that significant deviations of the Corrected Pad\'{e}  formula (\ref{fin4}) and of the Standard Pad\'{e} formula (\ref{form2}) from the reference rational expression (\ref{proxy}), start around $x=0.85$. All formulas start to depart from the original series around $x=0.8$.
The two formulae, (\ref{fin4}) and (\ref{form2}), happen to be very  close to each other almost everywhere, except in the immediate vicinity of the critical point.

\subsection{Accurate final formula}

According to our calculations, based on various re-summation techniques applied to the series (\ref{series}), we conclude that the critical amplitude is in the interval from $5.14$ to $5.24$, by $6-9\%$ higher than following naively to  O'Brien's  $4.82$. 

%Using instead of $\sqrt{3}$, another combination which includes purely geometrical factor $\sqrt{3} \sqrt{\frac{x_c}{x_c^{\text{sq}}}}$, where $x_c^{\text{sq}}$ is the threshold for the conductivity of a square lattice $x_c^{\text{sq}}=\frac{\pi }{4}$, one can obtain much better estimate for the amplitude $A=\frac{\sqrt[4]{3} \pi ^{3/2}}{\sqrt{2}}\approx5.18191$.

Below we present  an exceptionally accurate and more compact formula for the effective conductivity (\ref{form1}) valid for all concentrations.

Let us start with modified expression (\ref{and1}) taking into account also the O'Brien suggestion already in the starting approximation for the amplitude in  $\sigma^{r-l}$. All remaining parameters in (\ref{and1}) are obtained by matching it asymptotically with the truncated series $\sigma_{2}=1+2x+2x^2$,
\begin{equation}
\label{fin}
\begin{array}{llll}
\sigma_{2}^{r-l}(x)=
\frac{4.82231}{(0.9069\, -x)^{1/2}}-5.79784+
\\
2.13365 (0.9069\, -x)^{1/2}-0.328432 (0.9069\, -x).
\end{array}
\end{equation}

Repeating the procedure developed in subsection \ref{subs}, we receive several reasonable estimates for the critical amplitude, $A_7=5.18112$, $A_{11}=5.15534$, $A_{12}=5.19509$, $A_{13}=5.18766$. 

Complete expression for the effective conductivity corresponding to the first estimate for the amplitude, is given as follows

\begin{equation}
\label{fin1}
\sigma_{7}^{*}(x)=\sigma_{2}^{r-l}(x) F_{7}(x),
\end{equation} 
and
$F_{7}(x)=\frac{f_1(x)}{f_2(x)}$,
where
\begin{equation}
\label{fin2}
\begin{array}{llll}
f_{1}(x)=
52.0141 + 10.3198 x - 38.8957 x^2 + 4.70555 x^3 +\\
 4.89777 x^4 + 4.6887 x^5 + 0.476241 x^6 +7.49464 x^7,
\end{array}
\end{equation} 
and
\begin{equation}
\label{fin3}
\begin{array}{llll}
f_{2}(x)=
52.0141+10.3198 x-38.8957 x^2+2.17078 x^3+\\
5.80088 x^4+6.03946 x^5+1.80866 x^6+x^{7}.
\end{array}
\end{equation}

The formulae predict a sharp increase from $\sigma_{7}^{*}(0.906)=166.708$, to $\sigma_{7}^{*}(0.9068)=513.352$, in the immediate vicinity of the threshold, where other approaches \cite{McPh, hexa1,LB}, fail to  to produce an estimate. At the largest concentration $x=0.9068993$ mentioned in \cite{McPh}, the conductivity is very large, $8375.34$. %VM
This formula \eqref{fin1} after slight modifications can be written in the form \eqref{final}.

Asymptotic expression can be extracted from for the approximant (\ref{fin1}),
\begin{equation}
\label{as}
\sigma^{*}\simeq\frac{5.18112}{\sqrt{0.9069\, -x}}-6.229231.
\end{equation}

Even closer agreement with numerical results of \cite{McPh} is achieved with approximant corresponding to  $A_{13}$.
\begin{equation}
\label{fin4}
\sigma_{13}^{*}(x)=\sigma_{2}^{r-l}(x) F_{13}(x),
\end{equation} 
where $F_{13}(x)=\frac{f_1(x)}{f_3(x)}$,

\begin{equation}
\label{fin24}
\begin{array}{llll}
f_{1}(x)=
1.49313+1.30576 x+0.383574 x^2+0.467713 x^3+
\\
0.471121 x^4 + 0.510435 x^5 + 0.256682 x^6 +
\\
0.434917 x^7+0.813868 x^8+0.961464 x^9+
\\
0.317194 x^{10}+0.377055 x^{11}-1.2022 x^{12}-0.931575 x^{13};
\end{array}
\end{equation} 
and
\begin{equation}
\label{fin34}
\begin{array}{llll}
f_{3}(x)=
1.49313+1.30576 x+0.383574 x^2+0.394949 x^3+
\\
0.44785 x^4+0.503394 x^5+0.303285 x^6+
\\
0.271498 x^7+0.732764 x^8+0.827239 x^9+
\\
0.25509 x^{10}+0.239752 x^{11}-1.26489 x^{12}-x^{13}.
\end{array}
\end{equation}

It describes even more accurately than (\ref{fin1}), the numerical data in the interval from $x= 0.85$ up to the critical point. The maximum error for the formula (\ref{fin4}) is truly negligible, $-0.042\%$.

Asymptotic expression can be extracted from for the  approximant (\ref{fin4}),
\begin{equation}
\label{as1}
\sigma^{*}\simeq\frac{5.18766}{\sqrt{0.9069\, -x}}-6.2371.
\end{equation}
\subsubsection{Role of randomness}
For random two-dimensional composite we obtained recently \cite{0}, the following closed-form expression for the effective conductivity, 

\begin{equation}
\label{numer2}
\begin{array}{lll}
\sigma^{*}(x)=0.121708 f_{0,r}^{*}(x)
\times
\\
\exp \left(\frac{(0.64454 x-1.38151) x+0.72278}{(x-0.9069)^2 \sqrt{\frac{x (x+0.435329)+0.3582}{(x-0.9069)^2}}}-0.815613 \sinh ^{-1}\left(\frac{2.0171 (x+0.494058)}{x-0.9069}\right)\right), 
\end{array}
\end{equation}
with 
\begin{equation}
f_{0,r}^{*}(x)=\frac{(0.419645 x+1)^{3.45214}}{\sqrt{1-1.10266 x}}.
\end{equation}

Closed-form expression for the effective conductivity of the regular hexagonal array of disks is given by (\ref{fin4}).
Since the two expressions are defined in the same domain of concentrations, a comparison can explicitly quantify the role of a randomness (irregularity) of the composite.  In order to estimate an enhancement factor due to randomness we use ratio of (\ref{numer2}) to (\ref{fin4}). In particular, the enhancement factor at $x=0.906$, is equal to $104.593$.
 \begin{figure}
\begin{center}
\includegraphics[scale=0.7]{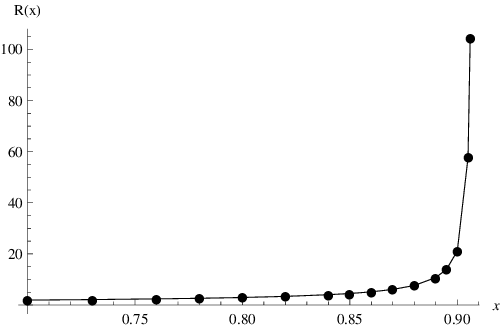}
\end{center}
\caption{
Ratio $R(x)=\frac{\sigma^*(x)}{\sigma^*_{13}(x)}$ of the effective conductivity for the random composite to the effective conductivity of the hexagonal regular lattice calculated with (\ref{numer2}) and (\ref{fin4}), respectively.
}
\label{figureenh}
\end{figure}
In Figure \ref{figureenh}, such an enhancement factor is shown in the region of high concentrations.

\section{Interpolation with High-concentration Pad\'{e} approximants
\label{high}
}
When two expansions (\ref{series})  and (\ref{as}) are available, the problem of reconstruction greatly simplifies and can be solved upfront in terms of Pad\'{e} approximants.

This approach requires as an input at least two parameters from weak and strong- coupling (high-concentration) regimes, including the value of amplitude $A=5.18112$ from (\ref{as}). Similar problem for random composites was considered in \cite{andperc}.

Assume that the next-order term, $B=-6.22923$ from (\ref{as}), is known in advance. The high-concentration limit, In terms of $z$-variable (\ref{direct}), the strong-coupling limit  is simply
\begin{equation}
\label{strong}
\sigma\simeq \frac{A}{\sqrt{x_c} }\sqrt{z}+B+O(z^{-1/2}).
\end{equation}
The Pad\'{e} approximants all conditioned to give a constant value as $z\rightarrow 0$ are given below,
\begin{equation}
\label{padenew}
\begin{array}{llll}
p_{2,1}(z)=\frac{\beta  \sqrt{z} \left(1+\beta _1 \frac{1}{\sqrt{z}}+\frac{\beta _2}{z}\right)}{1+\beta _3 \frac{1}{\sqrt{z}}},
\\
p_{3,2}(z)=\frac{\beta \sqrt{z} \left(1+\beta _1 \frac{1}{\sqrt{z}}+\frac{\beta _2}{z}+\beta _3 z^{-3/2}\right)}{1+\beta _5 \frac{1}{\sqrt{z}}+\frac{\beta _6}{z}},
\\
p_{4,3}(z)=\frac{\beta \sqrt{z} \left(1+\beta _1 \frac{1}{\sqrt{z}}+\frac{\beta _2}{z}+\beta _3 z^{-3/2}+\frac{\beta _4}{z^2}\right)}{1+\beta _5 \frac{1}{\sqrt{z}}+\frac{\beta _6}{z}+\beta _7 z^{-3/2}}.

\end{array}
\end{equation} 

The unknowns in (\ref{padenew}) will be obtained by the asymptotic conditioning to (\ref{strong}) and (\ref{series}).  In all orders $\beta=\frac{A}{\sqrt{x_c}}$. Explicitly, in original variables, the following expressions transpire,
\begin{equation}
\label{newexplicit}
\begin{array}{llll}
p_{2,1}(x)=\frac{\sqrt{\frac{x}{0.9069-x}}+\frac{4.9348}{0.9069-x}-4.11284}{\sqrt{\frac{x}{0.9069-x}}+1.32856},
\\
\\
p_{3,2}(x)=\frac{\left(0.608173 \sqrt{\frac{x}{0.9069-x}}+1.26563\right) x+0.677749 \sqrt{\frac{x}{0.9069-x}}+1.13282}{-\left(0.747325 \sqrt{\frac{x}{0.9069-x}}+1\right) x+0.677749 \sqrt{\frac{x}{0.9069-x}}+1.13282},
\\
\\
p_{4,3}(x)=\frac{5.4414 \sqrt{z(x)} \left(1+3.76414 \frac{1}{\sqrt{z(x)}}+\frac{7.73681}{z(x)}+1.97396 z(x)^{-3/2}+\frac{3.76815}{z(x)^2}\right)}{1+4.90893\frac{1}{\sqrt{z(x)}}+\frac{10.7411}{z(x)}+20.504 z(x)^{-3/2}}.\end{array}
\end{equation} 

The approximants are strictly non-negative and respect the structure of (\ref{series}), e.g. for small $x$, 
\begin{equation}
p_{4,3}(x)\simeq 1+2x+2 x^2+O(x^3),
\end{equation} 
since all lower-order powers generated by square-roots, are suppressed by design. But in higher order, emerging integer powers of  roots should be suppressed again and again, to make sure that only integer powers of $x$ are present. As $x\rightarrow x_c$,  
\begin{equation}
p_{4,3}(x)\simeq A (x_c-x)^{-1/2}+B +O((x_c-x)^{1/2}),
\end{equation} 
and only integer powers of a square-root appear in higher-orders. Both $p_{3,2}(x)$ and $p_{4,3}(x)$ give good estimates for the conductivity, from below and above respectively.  Their simple arithmetic average works better than each of the approximants. The bounds hold till the very core of the high-concentration regime, till $x=0.906$.

Particularly clear form is achieved for the resistivity, an inverse of conductivity, $r(z)=(p(z))^{-1}$, e.g.,

\begin{equation}
\label{resist}
r_{3,4}(z)=\frac{3.76815+1.97396 \sqrt{z}+0.902145 z+0.183776 z^{3/2}}{3.76815+1.97396 \sqrt{z}+7.73681 z+3.76414 z^{3/2}+z^2}.
\end{equation}

With the variable $X=\sqrt{z}$,  the resistivity problem is reduced to studying the sequence of Pad\'{e} approximants $R_{n}=r_{n,n+1}(X)$, $n=1,2...l/2$, with $X\in[0,\infty)$, and analogy with the Stieltjes truncated moment problem 
\cite{adam,trunc,kr}, 
is complete as long as the resistivity expands at $X\rightarrow\infty$ in the Laurent polynomial with the sign-alternating coefficients,  coinciding with the ``Stieltjes-moments" $\mu_{k}$ (see e.g., \cite{st,st1}, were the original work of Stieltjes is explained very clearly). 

The moments formally define corresponding Stieltjes integral as $X\rightarrow\infty$,
\begin{equation}
\label{expa}
\int _0^{\infty }\frac{d\phi(u) }{u+X}\sim\sum _{k=0}^l (-1)^k \mu _k X^{-k-1}+O( X^{-l}),
\end{equation}
$l$ is even \cite{trunc}, and $\mu_{k}=\int_0^{\infty } u^k \, d\phi(u)$. Approximant $R_{n}(X)$ should match (\ref{expa})
asymptotically.

The Stieltjes moment problem can possess a unique solution or multiple solutions, dependent on the behavior of the moments, in contrast with the problem of moments for the finite interval \cite{moment,moment1,moment2}, which is solved uniquely if the solution exists \cite{Wall}. The role of variable is played by the contrast parameter, while in our case of a high-contrast composite, the variable is $X$.

In our setup,there are just two moments available and resistivity is reconstructed using also a finite number of coefficients in the expansion at small $X$. I.e., the reduced (truncated) two point Pad\'{e} approximation is considered, also tightly related to the moment problem \cite{kr,mom1,mom2}. In fact, even pure interpolation problem can be presented as a moment problem.
We obtain here upper and lower bounds for resistivity (conductivity) in a good agreement with simulations \cite{McPh}. 

It does seem interesting and non-trivial that the effective resistivity (conductivity) can be presented in the form of a Stieltjes integral \cite{Wall,st,st1}, when the variable (\ref{direct}) is used.

\subsection{Independent estimation of the amplitude B
\label{indep}
}
We intend to calculate the amplitude B independent on previous estimates. 
Start with the choice of the simplest approximant as zero-approximation,
\begin{equation}
\label{padestart}
p_{1,0}(z)=\beta  \sqrt{z} \left(\frac{\frac{1}{\sqrt{z}}}{\beta }+1\right).
\end{equation} 

\begin{equation}
\label{padeor}
p_{1,0}(x)=5.4414 \sqrt{\frac{x}{0.9069\, -x}}+1,
\end{equation} 

\begin{figure}
\begin{center}
\includegraphics[scale=0.6]{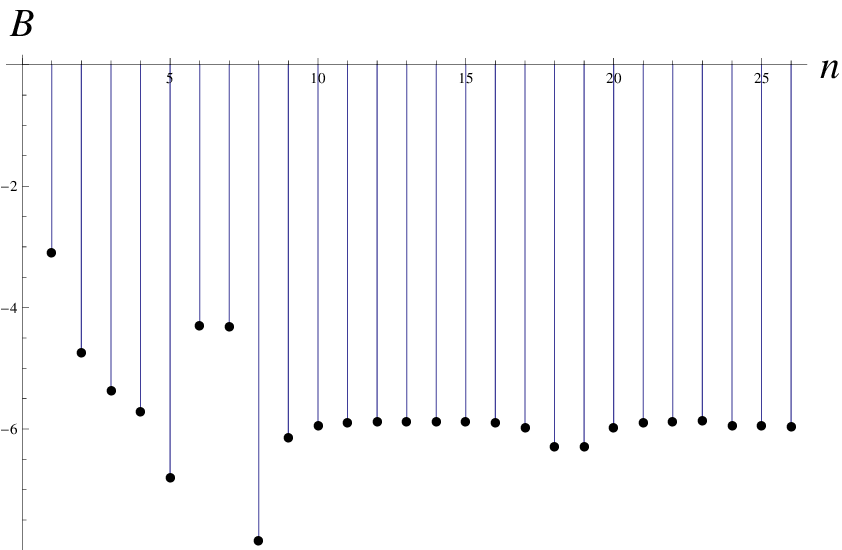}
\end{center}
\caption{
Sequence of approximations $B_{n}$ calculated from (\ref{amppadecor}).
}
\label{figureamp}
\end{figure}

The way how we proceeded above was to look for multiplicative corrections to some plausible "zero-order" approximate solution.  We can also look for an additive corrections in a similar fashion.  To this end subtract (\ref{padeor}) from (\ref{seriesreg}) to get some new series $g(x)$. Change the variable $x=y^2$ to bring the series to a standard form. The diagonal Pad\'{e} approximants to the series $g(y)$ are supposed to give a correction to the value of $1$, suggested by (\ref{padeor}). To calculate the correction one has to find the value of the corresponding approximant as  $y\rightarrow \sqrt{x_c}$.
The following sequence of approximations for the amplitude $B$ can be calculated now readily,
\begin{equation}
\label{amppadecor}
B_{n}=1+ PadeApproximant[g(y\rightarrow \sqrt{x_c}),n,n].
\end{equation}
The sequence of approximations is shown in Fig.\ref{figureamp}.

??? move $B$ in the $y$-axis 

There is clear saturation of the results for larger $n$, and $B_{26}=-5.94966$. 
One can reconstruct the expression for conductivity corresponding to $B_{26}$ in additive form
\begin{equation}
\label{highadd}
\sigma_{26}^{*}(x)=p_{1,0}(x)+F_{26}(x),
\end{equation} 
where $F_{26}(x)=\frac{F_2(x)}{F_6(x)}$,

\begin{equation}
\label{add26}
\begin{array}{llll}
F_{2}(x)=
-5.71388 \sqrt{x}-\\
1.5564 x-0.358877 x^{3/2}-2.18519 x^2+0.0918426 x^{5/2}-1.59468 x^3-
\\
0.149418 x^{7/2}-1.47691 x^4-0.366848 x^{9/2}-1.49733 x^5-0.56432 x^{11/2}-
\\
1.58738 x^6+0.21344 x^{13/2}-1.31081 x^7-0.366156 x^{15/2}+15.1037 x^8-
\\
15.1703 x^{17/2}-6.38227 x^9+0.576004 x^{19/2}-2.5147 x^{10}+0.715526 x^{21/2}-\\
1.53752 x^{11}+0.28655 x^{23/2}-1.19851 x^{12}+5.9511 x^{25/2}+0.558011 x^{13},
\end{array}
\end{equation} 
and
\begin{equation}
\label{add36}
\begin{array}{llll}
F_{6}(x)=
1+0.622415 \sqrt{x}-0.27066 x+0.294568 x^{3/2}-0.00182913 x^2+
\\
0.0875453 x^{5/2}+0.0832152 x^3+0.098912 x^{7/2}+0.114562 x^4+0.116471 x^{9/2}+
\\
0.133685 x^5+0.133737 x^{11/2}-0.0205003 x^6+0.0451001 x^{13/2}+
\\
0.134146 x^7-2.77482 x^{15/2}+1.6976 x^8+3.12806 x^{17/2}-0.87267 x^9+\\
0.16541 x^{19/2}-0.179645 x^{10}+0.0404152 x^{21/2}+0.000620289 x^{11}+\\
0.0514796 x^{23/2}-0.986857 x^{12}-0.58415 x^{25/2}+0.388415 x^{13}.
\end{array}
\end{equation}
 The maximum error for the formula (\ref{highadd}) is very small, $0.0824\%$, only slightly inferior compared with (\ref{fin4}). The amplitude B is firmly in the interval $(5.95,6.22)$, according to our best two formulae.

\section{Discussion of the ansatz (\ref{and},\ref{and1})
\label{ansatz}
}

In the case of a square lattice of inclusions \cite{McPh, square1, square2, square3, our1}, we looked for the solution in a simple form,

\begin{equation}
\label{square}
\sigma_{1}^{r-l}=\alpha_{1} (x_{c}-x)^{-1/2}+\alpha_{2}, \quad x_c=\frac{\pi}{4},
\end{equation}
and obtained the unknowns from the two starting terms of the corresponding series, 
\begin{equation}
\label{seriessquare}
\sigma\simeq 1+2x+2 x^2+2 x^3 +2 x^4+...
\end{equation}
Then, $\alpha_{1}=\frac{\pi ^{3/2}}{2}\approx2.784$, $\alpha_{2}=(1-\pi)$, same form as obtained asymptotically in \cite{MCass}, with exactly the same value for the leading amplitude as obtained in \cite{kell1}. 

Also the lower bound for amplitude $A$ is equal to $2.753$, and the upper bound is equal to $=2.856$, as can be found directly from the corresponding corrected Pad\'{e} sequences for the critical amplitude \cite{our1}.

Formula (\ref{square}) despite its asymptotic nature, turned out to be rather accurate in the whole region of concentrations. We try to understand below why it is so.

Let us subtract the approximant (\ref{square}) from the series (\ref{seriessquare}), apply to the new series transformation (\ref{direct}). Then we apply to such transformed series another procedure, intended to find corrections to the values of amplitudes $\alpha_{1}$ and $\alpha_{2}$. Such task is non-trivial, especially when one is interested in analytical solutions. It can be solved using general form of root approximants derived in \cite{cross,cross1}, 
\begin{equation}
\sigma_{add}=b_{0}  z^2 \left(\left(b_1 z+1\right){}^{s_1}+b_2 z^2\right){}^{s_2}
\end{equation}
under asymptotic condition
\begin{equation}
\sigma _{add}\simeq d_1 \sqrt{z}+d_2, \quad \mbox{as}\; z\rightarrow\infty.
\end{equation}

Elementary power-counting gives $s_1=3/2$, $s_2=-3/4$. All other unknowns can now  be determined uniquely in a standard fashion from the condition of asymptotic equivalence as $z\rightarrow0$.
Final expression
\begin{equation}
\sigma _{add}=\frac{0.0556033 x^2}{(0.785398\, -x)^2 \left(\frac{3.69302 x^2}{(0.785398\, -x)^2}+\left(\frac{1.98243 x}{0.785398\, -x}+1\right)^{3/2}\right)^{3/4}},
\end{equation}
can be re-expanded in the vicinity of $x_{c}$ with the result
\begin{equation}
\sigma _{add}\simeq \frac{0.0184973}{\sqrt{0.785398\, -x}} -0.0118315+O(\sqrt{x_c-x}),
\end{equation}
indicating only small corrections to the values of amplitudes.  Such asymptotic stability of all amplitudes additionally justifies the ansatz, and final corrected expression $\sigma^{sq}=\sigma_{1}^{r-l}+\sigma _{add}$, appears to be just slightly larger than (\ref{square}). Note that modified Pad\'{e} approximants as described above, are only able to produce additive corrections in the form $\sigma _{add}\simeq d \sqrt{z}+O(\frac{1}{\sqrt{z}})$ as $z\rightarrow\infty$.

In the case of hexagonal lattice, such simple proposition as $\ref{square}$ does not appear to be stable in the sense   described above. We have to try lengthier expressions of the same type, such as (\ref{and1}).  Additive correction in the form $\sigma_{add}=b_{0}  z^4 \left(b_2 z^2+\left(b_1 z+1\right){}^{3/2}\right){}^{-7/4}$, or
\begin{equation}
\sigma_{add}(x)=\frac{0.00220821 x^4}{(0.9069\, -x)^4 \left(\frac{21.8184 x^2}{(0.9069\, -x)^2}+\left(\frac{3.48493 x}{0.9069\, -x}+1\right)^{3/2}\right)^{7/4}},
\end{equation}
leads to the very small, almost negligible asymptotic corrections to the ansatz (\ref{and1}).  E.g., the leading amplitude changes to the value of $5.09925$. Such asymptotic stability of all amplitudes justifies the ansatz. Of course, it also appears to be reasonable when compared with the whole body of numerical results. The lower bound$=5.0925$, and the upper bound bound$=5.298$, can be found directly from the corresponding corrected Pad\'{e} sequences.

\section{Square and Hexagonal United
\label{united}
}
From the physical standpoint of there is no qualitative difference between the properties of hexagonal and square   lattice arrangements of inclusions. Therefore one might expect that a single expression exists for the effective conductivity of the two cases. 

Mathematically one is confronted with the following problem: for the functions of two variables $\sigma_{sq} (x, x_c^{sq}) $ and $\sigma_{hex} (x,x_c^{hex})$, to find the transformation or relation which connects the two functions. (Here $x_c^{hex}\equiv x_c$).

Assuming that the expressions for both lattices are different only with respect to lattice parameters simplifies the task, but is not necessary. The problem is really simplified due to similar leading asymptotic terms in the dilute and highly concentrated limits. On general grounds, one can expect that up to some simply behaving "correcting" function of a properly chosen non-dimensional concentration, the two functions are identical. Below we do not solve the problem from the first principles, but address it within the limits of some accurate approximate approach.

We intend  to express $\sigma_{sq}$ and $\sigma_{hex}$ in terms of the corresponding non-dimensional variables,
$Z_{sq}=\frac{x_{c}^{sq}-x}{x_{c}^{sq}}$ and  $Z_{hex}=\frac{x_{c}^{hex}-x}{x_{c}^{hex}}$, respectively. Each of the variables is in the range between $0$ and $1$.

Then, we formulate a new ansatz which turns to be good both for square and hexagonal lattices,
\begin{equation}
\label{self}
\sigma^{u}=\alpha_1 \frac{1}{\sqrt{x_c-x}} \left(\alpha_2 \sqrt{x_c-x}+1\right){}^u,
\end{equation}
where $u$ is a control parameter introduced by the self-similar renormalization \cite{32,4} applied to the asymptotic form (\ref{square}). One can obtain the unknowns from the three starting terms of the corresponding series, which happen to be identical for both lattices under investigation. 

Then, the method of (\ref{subs}), when the ansatz (\ref{self}) is corrected through application of the Pad\'{e} approximants, is applied. Emerging diagonal Pad\'{e}-sequences for critical amplitudes are convergent for both lattices and  good results are simultaneously achieved in the same order, employing $24$ terms from the corresponding expansions. 

We select from the emerging sequences only approximants which are also holomorphic functions. Not all approximants generated by the procedure are holomorphic. The holomorphy of diagonal Pad\'{e} approximants in a given domain implies their uniform convergence inside this domain (A.A. Gonchar, see \cite{gonch}).  

Corresponding corrective Pad\'{e} approximants, $Cor_{12}^{hex}$, $Cor_{12}^{sq}$, are given below.
 compatible with the formulae given below. For the hexagonal lattice,
\begin{equation}
\label{finhex}
\sigma_{hex}^{*}(Z)=\sigma^{c,hex}(Z) Cor_{12}^{hex}(Z),
\end{equation} 
and for the square lattice,
\begin{equation}
\label{finsq}
\sigma_{sq}^{*}(Z)=\sigma^{c,sq}(Z) Cor_{12}^{sq}(Z).
\end{equation} 
The initial approximation for the hexagonal lattice,
\begin{equation}
\label{selfhex}
\sigma^{c,hex}(Z)=\frac{4.5509 \left(1-0.637832 \sqrt{Z}\right)^{1.49198}}{\sqrt{Z}},
\end{equation}
and for the square lattice,
\begin{equation}
\label{selfsq}
\sigma^{c,sq}(Z)=\frac{3.29343 \left(1-0.659155 \sqrt{Z}\right)^{1.1074}}{\sqrt{Z}}.
\end{equation}

Correction term for the hexagonal lattice,

\begin{equation}
Cor_{10}^{hex}(Z)=\frac{cor_{1}^{hex}(Z)}{cor_{2}^{hex}(Z)},
\end{equation}

and for the square lattice,
\begin{equation}
Cor_{10}^{sq}(Z)=\frac{cor_{1}^{sq}(Z)}{cor_{2}^{sq}(Z)}.
\end{equation}
Numerators and denominators of these expressions are given by polynomials,
\begin{equation}
\label{cor1hex}
\begin{array}{llll}
cor_{1}^{hex}(Z)=
4.77682-2.70811 Z-20.9607 Z^2+139.454 Z^3-
\\
411.652 Z^4+752.321 Z^5-935.979 Z^6+824.735 Z^7-
520.883 Z^8+\\
232.778 Z^9-70.2807 Z^{10}+12.8747 Z^{11}-1.07872 Z^{12}
\end{array}
\end{equation} 

\begin{equation}
\label{cor2hex}
\begin{array}{llll}
cor_{2}^{hex}(Z)=4.07267+1.71405 Z-35.8279 Z^2+172.528 Z^3-\\
461.417 Z^4+801.75 Z^5-964.958 Z^6+828.769 Z^7-\\
512.137 Z^8+224.545 Z^9-66.7017 Z^{10}+12.0594 Z^{11}-Z^{12}
\end{array}
\end{equation}

\begin{equation}
\label{cor1sq}
\begin{array}{llll}
cor_{1}^{sq}(Z)=
10.0658-55.5091 Z+96.8247 Z^2-149.99 Z^3+\\227.446 Z^4-298.272 Z^5+346.9 Z^6-341.722 Z^7+258.688 Z^8-\\138.017 Z^9+47.9727 Z^{10}-9.69297 Z^{11}+0.860976 Z^{12}
\end{array}
\end{equation}

\begin{equation}
\label{cor2sq}
\begin{array}{llll}
cor_{2}^{sq}(Z)=10.3657-57.833 Z+105.045 Z^2-175.546 Z^3+\\
298.234 Z^4-438.138 Z^5+532.347 Z^6-508.883 Z^7+
\\
362.562 Z^8-182.102 Z^9+60.161 Z^{10}-11.6563 Z^{11}+
 Z^{12}
\end{array}
\end{equation} 

\begin{figure}
\begin{center}
\includegraphics[scale=0.8]{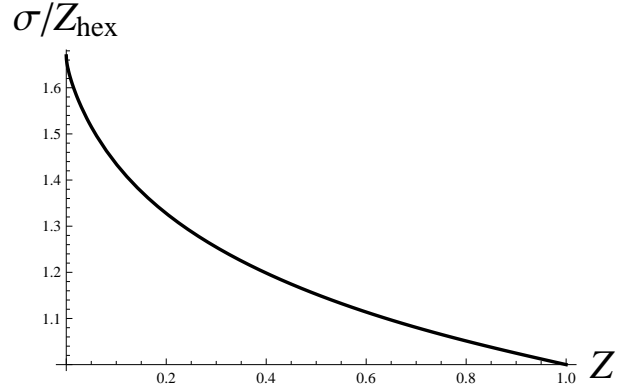}
\end{center}
\caption{
The ratio of final expressions for the conductivity of corresponding lattices,
$\frac{\sigma_{hex}^{*}(Z_{hex})}
{\sigma_{sq}^{*}(Z_{sq})}$, can be plotted (as $Z_{hex}=Z_{sq}=Z$).
}
\label{figuren}
\end{figure}

The ratio of final expressions for the conductivity of corresponding lattices,
$\frac{\sigma_{hex}^{*}(Z_{hex})}
{\sigma_{sq}^{*}(Z_{sq})}$, can be plotted (as $Z_{hex}=Z_{sq}=Z$), as shown in Fig. \ref{figuren}.

It  turns out that the ratio is bounded function of $Z$, and changes monotonously from $1$ ($Z=1$) to $1.669$ ($Z=0$).  The last number is not far from the  O'Brien suggestion \cite{McPh}, and is simply $\frac{A^{hex}}{A^{sq}}\sqrt{\frac{x_c^{sq}}{x_c^{hex}}}$.  Here $A^{hex}=5.08318$, $A^{sq}=2.834284$, are the critical amplitudes.

\section{Dirichlet summation. Large-$n$ behavior of series coefficients.
\label{Largen}
}

We will try to evaluate how the coefficients of the series behave at large-$n$.
From the practical viewpoint it is beneficial to have such information (if available), to be included into  resummation procedure. The so-called Borel summation is known to render filed-theoretical calculations more consistent.
With a similar goal, we employ the ordinary Dirichlet's series, defined conventionally $\phi (c)=\sum _{i=1}^{\infty} a_n n^{-c}$,
where $a_{n}$ stands for the coefficients of the original series.

Essential difference distinguishes the general theory of Dirichlet's series
from the simpler theory of power series. The region of convergence of a
power series is determined  by the position of the nearest singular points of the function which it represents.
The circle of convergence extends up to the nearest singular point. 
No such simple relation holds in the general case of
Dirichlet's series. When convergent in a portion of the plane they only may
represent a function regular all over the plane, or in a wider region of it. 

%The result is  that many of the peculiar
%difficulties which attend the study of power series on the circle of
%convergence are extended, in the case of Dirichlet's series, to wide
%regions of the plane or even to the whole of it.

However in an important case relevant to our study, the line of convergence
necessarily contains at least one singularity. It is covered by the following theorem:

Theorem 10 \cite{hardy}. If all the coefficients of the series are positive or
zero, then the real point of the line of convergence is a singular point of
the function represented by the series.

We conjecture, following \cite{complex}, that for large-$n$ the sum-function of coefficients, $S_n=a_1+a_2+...+a_n$,  behave as follows,
\begin{equation}
S_n\simeq \delta n^{c_1} \log ^{\varepsilon }(n).
\end{equation} 

Then, Dirichlet's series can be written explicitly in the form \cite{complex},
\begin{equation}
\phi (c)=\delta c \Gamma (\varepsilon +1) \left(c-c_1\right){}^{-\varepsilon -1}+g(c),
\end{equation}
where $g(c)$ stands for the regular part, and $\delta$ is a parameter. This expression is valid at $c>c_{1}$, where the Dirichlet's series are convergent.

In order to return to the physical region of variables $x$ and conductivity, let us apply the following transformation

\begin{equation}
c(x)=\frac{x_c (x+x_c)}{x_c-x},
\end{equation}

with the inverse 
\begin{equation}
x(c)=\frac{x_c c-{x_c}^2}{{x_c}+c},
\end{equation} 
with $c_1=x_c$,

The singular part of the conductivity after such transformation is expressed in the form
\begin{equation}
\sigma_{s}(x)=\frac{2^{-\varepsilon -1} \delta \Gamma (\varepsilon +1) (x+{x_c}) \left(\frac{x xc}{{x_c}-x}\right)^{-\varepsilon }}{x},
\end{equation}
and we should also set $\varepsilon=-1/2$. Parameter $\delta$ is simply connected with the critical amplitude $A$,
$\delta=\frac{A}{\sqrt{2 \pi } {x_c}}$.

Finally,
\begin{equation}
\sigma_{s}(x)=\frac{A \sqrt{\frac{x {x_c}}{{x_c}-x}} (x+{x_c})}{2 x {x_c}}.
\label{dirappp}
\end{equation}

This expression should also be regularized  at small $x$, so that 
\begin{equation}
\sigma_{s,r}(x)=\sigma_{s}(x)-\frac{A}{2 \sqrt{x}}.
\end{equation}

Close to critical point it can be expanded,

\begin{equation}
\sigma_{s,r}(x)\simeq \frac{A}{\sqrt{x_c-{x}}}-\frac{A}{2 \sqrt{{x_c}}}-\frac{A (-x+{x_c})}{4 {x_c}^{3/2}}+O((x_c-x)^{3/2}).
\end{equation}

After extracting the singular part from the series, the regular part expands for small $x$ into the following expression (only few low-order terms are shown)
\begin{equation}
g(x)\simeq 1-\frac{3 A \sqrt{x}}{4 {x_c}}+2 x-\frac{7 A x^{3/2}}{16 {x_c}^2}+2 x^2+O({x}^{5/2}),
\end{equation}
which is an expansion in $\sqrt{x}$.  To this expansion we apply  the diagonal  Pad\'{e} approximants.  Presence of fractional powers can be easily taken into account by change of variables, $x=y^2$, leading to doubling the number of approximants which can be constructed, compared with series of only integer powers.

E.g., in the lowest orders, in addition to a standard polynomial ratio with integer highest power, $\frac{-0.137912 x-3.89399 \sqrt{x}+1}{-0.460544 x+0.391414 \sqrt{x}+1}$, there is another ratio $\frac{-3.81871 \sqrt{x}+1}{0.4667 \sqrt{x}+1}$, with fractional highest power, which can be considered as a diagonal Pad\'{e} approximant too. Only the former-type polynomial ratios will be presented below, since the latter-type ratios do not bring better results in the current context.

Our goal now is to calculate the second, constant term in expansion close to $x_c$, denoted above as $B$. The correction to the constant term in the expansion emerges directly from the Pad\'{e} approximant calculated at $x=x_c$,

\begin{equation}
B_{n}=-\frac{A}{2 \sqrt{{x_c}}}+ PadeApproximant[g(x\rightarrow x_c),n,n].
\label{B}
\end{equation}

 We receive several reasonable estimates for the amplitude $B$: $B_{5}=-6.40157$, $B_{6}=-6.28506$, $B_{7}=-6.27028$, $B_{8}=-6.33762$, $B_{11}=-6.29595$, 
$B_{12}=-6.29695$, $B_{13}=-6.29842$. 

Explicitly in 7-th order, 
\begin{equation}
\sigma_{7}^{D}=\sigma_{s,r}(x)+PadeApproximant[g[x],7,7].
\end{equation}

Corresponding expressions for the singular part of solution,
\begin{equation}
\sigma_{s,r}(x)=\frac{\pi  \left(3 \sqrt{\frac{1}{\pi -2 \sqrt{3} x}} \left(2 \sqrt{3} x+\pi \right)-3 \sqrt{\pi }\right)}{2 \sqrt{2} 3^{3/4} \sqrt{x}},
\end{equation}

and for the regular part $g(x)=\frac{G_{1}(x)}{G_{2}(x)}$ given by the Pade approximant, we find

\begin{equation}
\label{G1}
\begin{array}{llll}
G_{1}(x)=
23.7835-88.5524 \sqrt{x}-39.6443 x+71.3743 x^{3/2}+\\
36.2957 x^2 - 12.3254 x^{5/2} - 5.54733 x^3 - 1.28303 x^{7/2} -\\
4.81208 x^4 - 1.00508 x^{9/2} - 4.36713 x^5 - 1.3826 x^{11/2} -\\
6.04028 x^6+4.92363 x^{13/2}+0.518137 x^7;

\end{array}
\end{equation}

\begin{equation}
\label{G2}
\begin{array}{llll}
G_{2}(x)=23.7835+13.3695 \sqrt{x}-29.9175 x-18.0154 x^{3/2}+\\
8.21262 x^2+6.49253 x^{5/2}+0.387922 x^3+1.2539 x^{7/2}+\\
0.50647 x^4+0.689113 x^{9/2}+0.505114 x^5+0.554584 x^{11/2}-\\
1.31685 x^6-0.498839 x^{13/2}+x^7.
\end{array}
\end{equation} 

The maximum error for the formula is small, just $-0.1602\%$.

The formulae predict the following values, $\sigma_{7}^{D}(0.906)=166.494$, $\sigma_{7}^{D}(0.9068)=512.7472$, $\sigma_{7}^{D}(0.9068993)=8376.58$. These value are very close to the predictions already presented above.

We conclude that our conjecture concerning the large-$n$ behavior of the sum-function of the coefficients, is in a good agreement with available numerical data. Also the estimates for $B$, which stem from the conjecture, is close to other estimates from the present paper.

Algorithms and mathematical methods used above, are based on asymptotic power-series for the effective conductivity  and various resummation techniques to ensure their convergence. Such approach is typical for Computational Science of Composite Materials. It is interesting to compare such approach and classic methodology based on direct solutions of  PDE's.

\section{Application of Lubrication Theory
\label{lubrication}
}
To find the effective conductivity in a classic way, one has to consider the local problem for Laplace equation describing regular hexagonal lattice of cylindrical inclusions. Such a study can be based on the Lubrication theory \cite{christ}, applicable for an asymptotic regime of large, ideally conducting inclusions. It has to be applied in conjunction with some averaging technique to derive effective conductivity. It is expedient first to consider inclusions with finite conductivity $\lambda$, and then to consider the limit $\lambda \to \infty$.
\begin{figure}
\begin{center}
\includegraphics[scale=0.6]{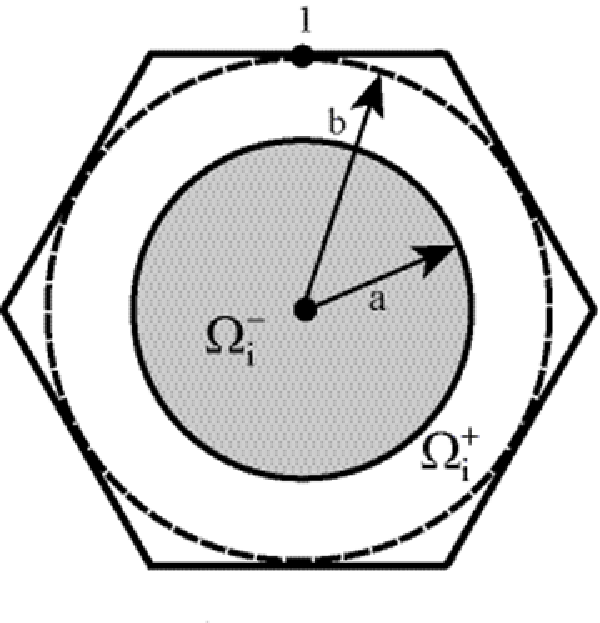}
\end{center}
\caption{
The hexagonal cell with the disk of the radius $a$ is approximated by the circle cell of the radius $b$.
}
\label{FigHex1}
\end{figure}

Main idea of the Lubrication theory consists in replacing the original boundary problem with another, corresponding to a simpler geometry (see Fig.\ref{FigHex1}). I.e. the original hexagonal elementary cell is replaced by a circle off radii $b$. Using so-called "fast" variables $(\xi,\eta)$ and the corresponding local polar coordinates $(r,\theta)$ we arrive at the following problem (for details see \cite{square1, square2})
\begin{equation}
\label{eq:lub1}
\frac{\partial^2 u}{\partial r^2} + \frac{1}{r} \frac{\partial u}{\partial r^2} +\frac{1}{r^2}\frac{\partial^2 u}{\partial \theta^2} =0, \quad r<a,\; a<r<b, 
\end{equation}
\begin{equation}
\label{eq:lub2}
u^+=u^-, \;\frac{\partial u^+}{\partial r} -\lambda \frac{\partial u^-}{\partial r}= (\lambda-1)(\cos \theta +\sin \theta), \quad r=a,
\end{equation}
\begin{equation}
\label{eq:lub3}
u =0,  \quad r=b,
\end{equation}
where $a$ is the radius of inclusions. For definiteness, the external flux is taken in such a way that the macroscopic flow is presented by the potential $u_0=(x_1,x_2)$ and the flux by the vector $(1,1)$ (for details see \cite{square1}). The problem \eqref{eq:lub1}-\eqref{eq:lub3} has the solution 
\begin{equation}
\label{eq:lub4}
u = \left\{
\begin{array}{llll}
N_1 r \cos \theta + N_2 r \sin \theta,  \quad r \leq a,
\\
(M_1 r+ \frac{K_1}{r})  \cos \theta + (M_2 r+ \frac{K_2}{r}) \sin \theta, \quad
a \leq  r \leq b,
\end{array}
\right.
\end{equation}
where the constants are determined by the boundary conditions 
\begin{equation}
\label{eq:lub5}
\begin{array}{llll}
N_1=N_2=\frac{(\lambda-1)(b^2-a^2)}{[b^2+a^2-\lambda(b^2-a^2)]},
\\
M_1=M_2=-\frac{(\lambda-1)a^2}{[b^2+a^2-\lambda(b^2-a^2)]},
\\
K_1=K_2=-\frac{(\lambda-1)a^2b^2}{[b^2+a^2-\lambda(b^2-a^2)]}.
\end{array}
\end{equation}

According to the Lubrication approach \cite{{christ}, square3}, let us consider an external contour for the cell, as a circle of varying radii  
\begin{equation}
\label{eq:lub6}
b(\xi)= \left\{
\begin{array}{llll}
2\sqrt{\xi^2-\sqrt{3} \xi +1},  \quad 0 \leq \theta < \frac{\pi}3,
\\
\sqrt{\xi^2 +1},  \quad \frac{\pi}3 \leq \theta \leq \frac{\pi}2,
\end{array}
\right.
\end{equation}

\begin{figure}
\begin{center}
\includegraphics[scale=0.8]{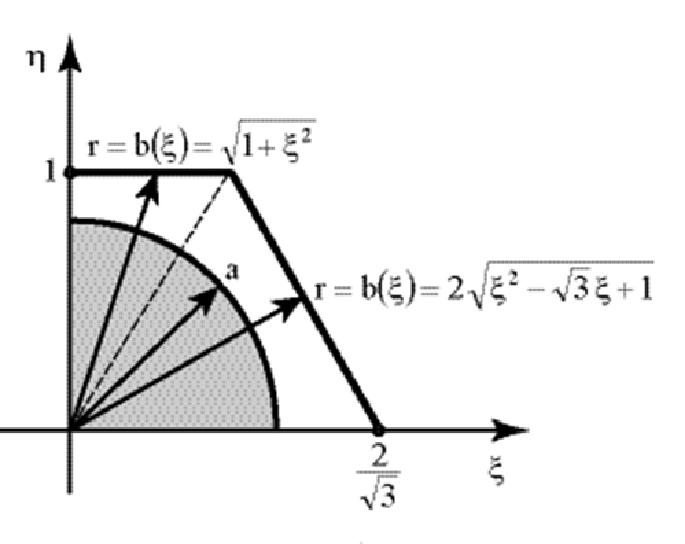}
\end{center}
\caption{
Approximation of the hexagonal cell by the circle cell of the variable radius $b(\xi)$.
}
\label{FigHex2}
\end{figure}

Integration is conducted over the quarter of the elementary cell, shown in Fig.\ref{FigHex2}. Following general prescriptions of the averaging method, we derive  averaged coefficient
\begin{equation}
\label{eq:lub7}
\sigma =\frac{1}{|\Omega|} \left[ \int_{\Omega_i^+} \left(1+\frac{\partial u}{\partial \xi}+\frac{\partial u}{\partial \eta} \right) d\xi d\eta + \lambda\int_{\Omega_i^-} \left(1+\frac{\partial u}{\partial \xi}+\frac{\partial u}{\partial \eta} \right) d\xi d\eta\right], 
\end{equation}
where $|\Omega|=2\sqrt{3}$. The integration is performed to satisfy also the relation \eqref{eq:lub6}; in particular, $b(\xi)$ is considered as a corresponding functions of varying radius. 

\subsection{Lubrication Approximation and Correction}
After some transformations we receive the following expression for the effective conductivity (or thermal conductivity) as the function of the inclusion size $a$,
\begin{equation}
\label{Lub}
\begin{array}{llll}
\sigma(a)=\frac{\left(2 \sqrt{3} a^2\right) \tan ^{-1}\left(\frac{\sqrt{3}}{3 \sqrt{1-a^2}}\right)}{\sqrt{1-a^2}}+1+\\
\frac{1}{3} \left(\sqrt{3} a^2\right) \left(\frac{\pi }{4}-\frac{3}{2} \sin ^{-1}\left(\frac{\sqrt{3}}{3 a}\right)\right)+
\frac{4 \sqrt{3} a^2}{3 \sqrt{1-a^2}}\times\\
\left(\tan ^{-1}\left(\frac{\left(\sqrt{3} a-\sqrt{3 a^2-1}\right) \sqrt{1-a^2}}{a+1}\right)-\right.\\
\frac{1}{4} \tan ^{-1}\left(\frac{2 \left(-\sqrt{3 a^2-1}+\sqrt{3} a-1\right) \sqrt{1-a^2}}{\sqrt{3 a^2-1} \left(\sqrt{3} a+a-2\right)+\left(1+\sqrt{3}\right) a \left(1-\sqrt{3} a\right)+2}\right)-\\
\left.\frac{1}{8} \tan ^{-1}\left(\frac{2 \sqrt{1-a^2}}{a}\right)\right)-\frac{1}{4} a^2 \log \left(\frac{3 a^2+2 \sqrt{3 \left(3 a^2-1\right)}+2}{4-3 a^2}\right)
\end{array}
\end{equation} 
As the inclusion size tend to its limiting value, $a\rightarrow 1$, the leading term in the conductivity of the ideally conducting inclusions can be found in the familiar form,
\begin{equation}
\label{Lub1}
\sigma_{0}\simeq \frac{\sqrt{\frac{3}{2}} \pi }{\sqrt{1-a}}.
\end{equation}
When expressed in terms of volume fraction of inclusions (\ref{Lub1}), coincides with Keller's formula \eqref{kellcrit}.

\begin{figure}
\begin{center}
\includegraphics[scale=1.]{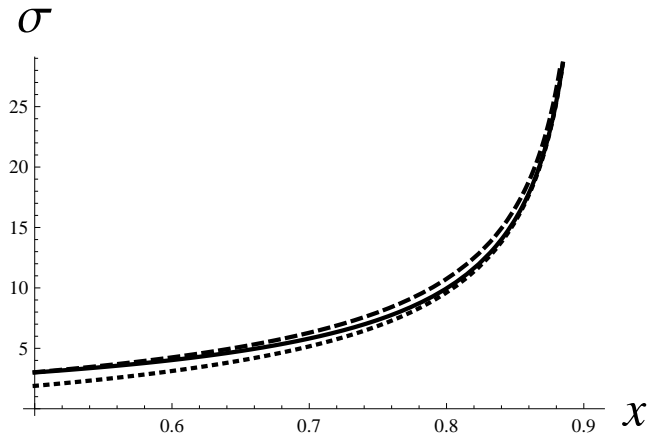}
\end{center}
\caption{
$\sigma$ calculated by formulas \eqref{fin24} (solid line), by \eqref{as1} (dotted line) and by \eqref{Lub3} (dashed line).  
}
\label{FigHex3}
\end{figure}

The first (constant) correction term to the formulae \eqref{Lub}, can be also obtained, leading to "shifted" expression for the conductivity in the critical region,
\begin{equation}
\label{Lub3}
\sigma_{1} \simeq \sigma_{0}-5.10217.
\end{equation}
Formula (\ref{Lub}) works rather well, with accuracy less than $2\%$, for concentrations as low as $x\simeq 0.82$. Its predictions for concentrations very close to $x_{c}$, are also very near to predictions from other formulae given above (see Fig.\ref{FigHex3}). Formula (\ref{Lub}) becomes invalid for $x\leq 0.3023$.

In the case of a square lattice of inclusions, Lubrication theory gives the following asymptotic result \cite{AAS},
\begin{equation}
\sigma{(x)}\simeq \frac{\pi^{3/2}}{2 \sqrt{\frac{\pi }{4}-x}}-1. \label{sq1}
\end{equation}

Formula \ref{sq1} should be compared with the more accurate result of \cite{MCass},\cite{MCass1},
\begin{equation}
\sigma{(x)}\simeq \frac{\pi^{3/2}}{2 \sqrt{\frac{\pi }{4}-x}}-\pi+1. \label{sq2}
\end{equation}

It appears that Lubrication theory assumptions, concerning reduction of the elementary cell to a circle, work better for the hexagonal lattice than for the square lattice. In both cases the correction term is overestimated. 

Classic approach to PDE's solution thus is limited to high-concentration asymptotic regime with strong interactions between inclusions.
 
On the other side, the whole well-developed family of self consistent methods which include Maxwell's approach, effective medium approximations, differential schemes etc., are valid only for a  dilute composites when interactions between inclusions do not matter \cite{Mit2013b}. 

In contrast, computational methods of the present paper are applicable everywhere.

Let us derive an interpolation formula by matching the two limiting expressions, (\ref{series}) and (\ref{Lub3}).
The method of sewing the two limiting behaviors together will be chosen to employ the main idea of Section~\ref{Largen}.
First we assume that the high-concentration formula (\ref{Lub3}) holds everywhere and then derive an additive correction in the form of the diagonal  Pad\'{e}  approximants in such a way that also the low-concentration limit (\ref{series}) is respected. It turns out that such approach not only generates another good interpolation formula, but also calculates an additive correction to the amplitude $B$. Technically, one should only replace the expression (\ref{dirappp}) with (\ref{Lub3}) and extract it from the \ref{series}, leading to the new series $g(x)$ and to 
corresponding approximations to the sought amplitude,
\begin{equation}
B_{n}=-5.10217+ PadeApproximant[g(x\rightarrow x_c),n,n].
\label{B1}
\end{equation}

We receive several reasonable estimates for the amplitude $B$:
$B_{5}=-6.37811$, $B_{6}=-6.29179$, $B_{7}=-6.28019$, $B_{8}=-6.42952$, $B_{11}=-6.29702$, 
$B_{12}=-6.29908$, $B_{13}=-6.32249$. These results are only slightly higher than estimates obtained above in Section~\ref{Largen}. Interpolation formula corresponding to $B_{7}$ is as accurate as its counterpart suggested in Section~\ref{Largen}.

\section{Random composite from hexagonal representative cell
\label{RAND}}
In the present paper, the numerical computations for random composites are performed for the hexagonal representative cell.  The number of inclusions per cell can be taken arbitrary large, hence the shape of the cell does somewhat influences the final result.

The hexagonal lattice serves as the domain Q,  where random composite is generated as a probabilistic distribution of disks of radius $r$ (particles), by means of some Monte-Carlo algorithm (protocol) \cite{method}.

%We consider a uniform non-overlapping distribution when    
%a set of independent and identically distributed (i.i.d.) points $\mathbf a_i$ are located in the plane in such a way that $|\mathbf a_i - \mathbf a_j| \geq 2r$. 

Algorithm 1, random sequential addition (RSA).
First random point  is randomly distributed in Q.  Second point is randomly distributed in Q with exception of the small circular region of radius $r$ surrounding the first point. Hence, the distribution of the second random
point is conditional and depends on the first random point.  More points, up to some number N, can be generated, conditioned that circular regions around all previous points are excluded from Q. This joint random variable for all points correctly determines sought probabilistic distribution. But the computer simulations work only up to concentrations as high as 0.5773, hence is the main RSA limitation. To overcome the limitation and to penetrate the region of larger concentrations, one has to apply some extrapolation technique.
%generation can be repeated N times.

Algorithm 2, random walks (RW) employed  also in \cite{0}.
N-random points are generated, at first being put onto the nodes of the hexagonal array. Let each point move in a randomly chosen direction with some step. Thus each center obtains new complex coordinate. This move is repeated many times, without particles overlap. If particle does overlap with some previously generated, it remains blocked at this step. After a large number of walks the obtained locations of the centers can be
considered as a sought statistical realization, defining random composite. 

RW protocol can be applied for arbitrary concentrations including those very close to $x_{c}$, which stands also for the maximum volume fraction of random composites. At $x=x{c}= \frac{\pi}{\sqrt{12}}$, we arrive to the regular hexagonal array of disks.

 The effective conductivity of random composite is also expected to tend to infinity as a power-law, as the concentration $x$ tends to the maximal value $x_{c}$,
\begin{equation}
\label{critr}
\sigma(x)\simeq A (x_c-x)^{-s} .
\end{equation}

The superconductivity critical exponent $s$ believed to be close to $\frac 43 \approx 1.3$ \cite{torkbook}, much different from the regular case. The critical amplitude $A$ is an unknown non-universal parameter.
We demonstrate below that $s$ depends on the protocol, and suggest simple way to decrease the dependence on protocol. Still, more studies are needed with different protocols.

Algorithm 2 allows to obtain the following power series in concentration,
\begin{equation}
\label{RW}
\sigma^{RW}=1+2 x+2 x^2+4.23721 x^3+6.8975 x^4.
\end{equation}
The higher order polynomial representations fail to give a non-zero value for the fourth-order coefficient.

Reasonable estimate for the critical index $s$ can be obtained already from the $D-Log$ formula combined with the transformation (\ref{direct}), (\ref{seq1}). 
Namely, the  result is $s_{2}= 1.43811$, and for the amplitude we obtain $A=1.21973$.
 
 The algorithm 1 produced the following series in concentration \cite{0},
\begin{equation}
\label{RSA}
\sigma^{RSA}=1+2 x+2 x^2+5.00392 x^3+6.3495 x^4.
\end{equation}
The coefficients on $x^k$ ($k=5,6,7,8$) vanish in \eqref{RSA} with the precision $10^{-10}$.

Good estimate for the critical index $s$ can be obtained already from the $D-Log$ formula (\ref{seq1}). The results are $s_{2}= 1.28522$ for the critical index, and  $A=1.57678$ for the amplitude.

Ideally, we would like to have $s$ and $A$ to be evaluated independent on protocol, but can hope only that combining two different protocols can decrease the dependence of $s$ on protocols, because errors of the two protocols can compensate. 

Assume that both schemes should lead to the same index, amplitude and threshold. Let us form a simple product, 
\begin{equation}
\label{J}
\sigma^{J}=\sqrt{\sigma^{RW} \sigma^{RSA}}.
\end{equation}

\subsection{D-Log estimates}
Apply now the $D-Log$ technique combined with the transformation (\ref{direct}), to the series (\ref{J}). The result is $s=s_{2}=1.34715$,  better than for each of the individual  components.

Simple addition of (\ref{RSA}) and (\ref{RW}) also leads to a good estimate $1.34888$, by the $D-Log$ technique.

Slightly better result is achieved for the geometrical mean of the series,
\begin{equation}
\label{M}
\sigma^{M}=\frac{2\sigma^{RW} \sigma^{RSA}}{\sigma^{RW} +\sigma^{RSA} },
\end{equation}
and $s=s_{2}=1.34542$. The coefficients in the expansion for small $x$,

\begin{equation}
\label{serM}
\sigma^{M}\simeq 1+2 x+2 x^2+4.62056 x^3+6.6235 x^4,
\end{equation} 
are formed as a compromise between the two algorithms. 

The effective conductivity can be reconstructed \cite{puller,padeour,0}, from an effective critical index (or $\beta$-function).
After some calculations, we obtain
\begin{equation}
\label{numM}
\begin{array}{lll}
\sigma^{M}_{*}(x)=3.24319 e^{0.441389 \tan ^{-1}\left(2.18756\, +\frac{2.43087}{x-0.9069}\right)}
\times
\\
\left(\frac{0.9069}{0.9069-x}-0.515166\right)^{1.33609} \left(\frac{x (x+0.0245056)+0.176696}{(0.9069\, - x)^2}\right)^{0.00466513}
\end{array}
\end{equation}
Also, the critical amplitude evaluates as $1.423$. Eq. (\ref{numM}) works as good as any other formula for the effective conductivity obtained in \cite{0}.

\subsection{"Single pole" approximation}
Critical index can be estimated also from a standard representation for the derivative

\begin{equation}
\label{Sp}
B_a(x)=\partial_x\log(\sigma^{M}(x))
\simeq \frac{s}{x_c-x},
\end{equation}
as $x\rightarrow x_{c}$, thus defining critical index as the residue in the corresponding single pole.

Outside of the immediate vicinity of the critical point a diagonal  Pad\'{e} approximant is assumed for the residue estimation\cite{pade1}, but such approach fails in the case under study. Let us use another representation, in the form of a factor approximant \cite{puller},
\begin{equation}
\label{Ba}
B_a(x)=\frac{2 (b_2 x+1)^{s_2}}{1-\frac{x}{x_c}},
\end{equation}
with the following values for parameters $b_2=7.84091$, $s_2=-0.140629$, found for the series (\ref{M}).

Formula (\ref{Ba}) leads to the simple expression for the critical index
\begin{equation}
s=2 x_c (b_2 x_c+1)^{s_2},
\end{equation}
and to the value $s=1.35129$.
The effective conductivity can be reconstructed as follows,
\begin{equation}
\label{numH}
\begin{array}{lll}
\sigma^{*}(x)=
\\
\exp \left(\frac{2 \pi  \left((b_2 x+1)^{s_2+1} \, _2F_1\left(1,s_2+1;s_2+2;\frac{2 \sqrt{3} (b_2 x+1)}{\pi  b_2+2 \sqrt{3}}\right)-
\, _2F_1\left(1,s_2+1;s_2+2;\frac{2 \sqrt{3}}{\pi  b_2+2 \sqrt{3}}\right)\right)}{\left(\pi  b_2+2 \sqrt{3}\right) (s_2+1)}\right),
\end{array}
\end{equation}
through the hypergeometric function. The "single pole" approximation (\ref{Sp}) is in fact equivalent to the particular case of the hypergeometric function.

For the RSA-series (\ref{RSA}), the same approach gives $s=1.31786$, while for the RW-series (\ref{RW}), $s=1.37978$. The difference between the two algorithms is small compared to all others methods employed for the index estimations.

\subsection{Corrected Index. Scheme 1}
We follow below the general idea of \cite{our1,0}, also explained in Subsection 
\ref{shem1}. At first, one should  obtain  an approximate solution explicitly as a factor approximant \cite{fac1,fac2}. Then we attempt to correct the form of the initial approximation with additional factor, originated from the part of series which did not participate in the formation of the initial approximation, following literally the way  leading to (\ref{betaf}).  

The simplest factor approximant can be calculated,
\begin{equation} 
\label{regf}
f_{0}(x)=\frac{(x+1)^{1.04882}}{(1-1.10266 x)^{0.862622}}.
\end{equation} 
 Such approximant satisfy the two non-trivial starting terms from the series (\ref{serM}), and incorporates the accepted value of the threshold $x_c$. It predicts for the critical index the value $s_{0}=0.862622$.
 
In the next step we attempt to correct $s_{0}$ using the $D-Log$-correction approach \cite{our1,0}, as described also in Subsection \ref{shem1}. Let us form the following ratio, $\frac{\sigma^{M}}{f_{0}(x)}$. Repeating the same steps that lead to the corrected expression for the index \ref{corindex}, we obtain the corrected value $s_{2}=1.32067$. Corresponding amplitude is equal to $1.48267$.
 
 The conductivity can be reconstructed in a closed form. Calculating corresponding integral
with $\beta$-function\cite{padeour,puller} $P_{2,3}(z)$, 
\begin{equation}
P_{2,3}(z)=\frac{5.71085 z^2}{12.4679 z^3+10.3351 z^2+4.31945 z+1},
\end{equation}
 we obtain
 \begin{equation}
\label{numer0}
\begin{array}{lll}
\sigma_{2}^{*}(x)=1.77719 \frac{(x+1)^{1.04882}}{(1-1.10266 x)^{0.862622}} e^{-0.465101 \tan ^{-1}\left(\frac{2.16258 x+0.451103}{0.9069-x}\right)}
\\
\times 
\left(\frac{0.545059 x+0.412586}{0.9069\, -x}\right)^{0.444818} \left(\frac{x^2+0.0241854 x+0.18073}{(0.9069-x)^2}\right)^{0.00661298}.
\end{array}
\end{equation}
\subsection{Corrected Index. Scheme 2}
Let us start from the initial approximation (\ref{regf}), and recast it more generally as 
\begin{equation}
f_{0}(x)=(1-\frac{x}{x_{c}})^{-s_{0}} R(x),
\end{equation}
where $R(x)$ stands for the regular part of (\ref{regf}).  In what follows we attempt to correct $f_{0}(x)$ differently than above, assuming instead of $s_{0}$ some functional dependence $S(x)$.

As $x\rightarrow x_{c}$, $S(x)\rightarrow s_{c}$, the corrected value. The function $S(x)$ will be designed in such a way, that it smoothly interpolates between the initial value $s_{0}$ and the sought value $s_{c}$. The corrected functional form for the conductivity  is now
\begin{equation}
\label{shem2}
f^{*}(x)=(1-\frac{x}{x_{c}})^{-S(x)} R(x).
\end{equation}
From (\ref{shem2}) one can express $S(x)$, but only formally since $f^{*}(x)$ is not known. But we can use its asymptotic form (\ref{serM}), express $S(x)$ as a series and apply some resummation procedure (e.g. Pad\'{e} technique). Finally calculate the limit of the approximants as $x\rightarrow x_{c}$. 

In what follows the ratio $C(x)=\frac{\sigma^{M}(x)}{R(x)}$, stands for an asymptotic form of the singular part of the solution, and as $x \rightarrow 0$
\begin{equation}
S(x)\simeq \frac{\log \left(C(x)\right)}{\log \left(1-\frac{x}{x_{c}}\right)},
\end{equation}
which can be easily expanded in powers $x$,  around the value of $s_{0}$. It appears that one can construct a single  meaningful Pad\'{e} approximant,
\begin{equation}
S(x)=\frac{4.91072 x^2+0.703479 x+0.862622}{3.00966 x^2+0.815512 x+1},
\end{equation}
and find the corrected index, $s_{c}=S(x_{c})=1.31426$. Now we also possess a complete expression for conductivity (\ref{shem2}). 

Scheme 2 due to its simplicity, can always lead to the analytical expression. But Scheme 1 seems to be the most flexible. It also turns out to be weakly dependent on the starting approximation $f_{0}(x)$. Indeed, if another starting approximation is considered,
\begin{equation}
f_{0}(x)=\frac{(2x+1)^{0.355391}}{(1 -1.10266 x)^{1.16919}},
\end{equation}
the corrected index remains good, $s_{3}=1.31094$.

The conductivity again can be reconstructed in a closed form. Calculating corresponding integral
with $\beta$-function $P_{3,4}(z)$, 
\begin{equation}
P_{3,4}(z)=\frac{9.85652 z^3+4.06592 z^2}{69.5337 z^4+35.9326 z^3+20.6483 z^2+6.17673 z+1},
\end{equation}
 we obtain rather lengthy expression,
\begin{equation}
\label{numer1}
\begin{array}{lll}
\sigma_{3}^{*}(x)=1.05615 \frac{(2x+1)^{0.355391}}{(1 -1.10266 x)^{1.16919}} 
\times
\\
\exp \left(0.0615685 \tan ^{-1}\left(2.23114\, -\frac{2.10073}{0.9069\, -x}\right)-0.0789727 \tan ^{-1}\left(4.693\, -\frac{5.46758}{0.9069\, -x}\right)\right)
\\
\times \left(\frac{x (1-0.245701)+0.138583}{(0.9069\, -x)^2}\right)^{0.081481}/\left(\frac{x ( x+0.415099)+0.099469}{(0.9069\, -x)^2}\right)^{0.0106051}.
\end{array}
\end{equation}
The form of expressions (\ref{numer0}), (\ref{numer1}) is unlikely to be guessed as an independent approximant.
  
\section{Conclusion
\label{discussion}
}

Based on estimates for the critical amplitudes $A$ and $B$, we derived an accurate and relatively compact formula for the effective conductivity (\ref{fin1}) valid for all concentrations, including the most interesting regime of very high concentrations. For the high-concentration limit, in addition to the amplitude value of $5.18112$, we deduce also that the next order (constant) term $B$, equals $-6.22923$. It is possible to extract more coefficients in the high-concentration expansion based on the formula (\ref{self}). Dirichlet summation is suggested to extract an arbitrary large-$n$ behavior of the coefficients.

When two expansions around different points (\ref{series})  and (\ref{as}) are available, the problem of reconstruction can be solved in terms of high-concentration Pad\'{e} approximants, implying that the effective resistivity (conductivity) can be presented in the form of a Stieltjes integral, 
%$\int _0^{\infty }\frac{d\phi(u) }{u+X}$,
%where $\phi(u)$ is a bounded non-decreasing function of $u$ \cite{st,Wall}, 
in terms of the variable $X=\sqrt{\frac{x}{x_c-x}}$.
Such Pad\'{e} approximants give tight lower and upper bounds for the conductivity, valid 
up to the very high $x$.

Such properties as the superconductivity critical index and threshold for conductivity, can be calculated from the series (\ref{series}). In the case of truncated series, the standard  Pad\'{e} approximants are not able to describe the correct asymptotic behavior in the high-concentration limit, where in addition to the leading critical exponent also a non-trivial sub-leading exponent(s) plays the role \cite{cross,cross1}. On the other hand when such a non-trivial  asymptotic behavior is treated separately with different type of approximants, the Pad\'{e} approximants are able to account for the correction. Such patchwork approximations appear to be more accurate and powerful than  approximating conventionally with a single type of approximants.

Simple functional relation between the effective conductivity of the hexagonal and square lattices is suggested, expressed in terms of some bounded monotonous function of a non-dimensional concentration of inclusions. Getting an accurate formula in this case, means that correct asymptotic behavior (\ref{as}) is indeed can be extracted from the series (\ref{series}), and together they determine the behavior in the whole interval with good accuracy. Neglecting the high-concentration regime dominated by necks, is not admissible.

We also considered a classic approach  based on Lubrication theory and concluded  that it can be applied strictly within the high-concentration asymptotic regime. In contrary, the celebrated  Maxwell's approach, effective medium approximations and differential schemes are valid only for a  dilute composites \cite{Mit2013b}. Computational approach and results of the present paper are applicable everywhere.

We conclude that approach based on the long power series for the effective conductivity as a function of particle volume fraction can be consistently applied in the important case of highly conducting (superconducting) inclusions. Based on our investigation we put forward the final formula \eqref{final}, for the effective conductivity of the hexagonal array.

\begin{acknowledgement}
The authors are grateful to Leonid Berlyand for stimulating discussion.
\end{acknowledgement}


\begin{thebibliography}{99}

\bibitem{adam}
V. M. Adamyan,  I. M. Tkachenko and M. Urrea, Solution of the Stieltjes Truncated Moment Problem,  Journal of Applied Analysis {\bf 9},  57-74 2003

\bibitem{andperc}
I. V. Andrianov, V.V. Danishevskyy, A. L. Kalamkarov, Analysis of the effective conductivity of composite materials in the entire range of volume fractions of inclusions up to the percolation threshold, Composites: Part B {\bf 41}, 503 -507, 2010

\bibitem{AAS}
I. V. Andrianov, J. Awrejcewicz, G. A. Starushenko. Application of an improved three-phase model to calculate effective characteristics for a composite with cylindrical inclusions. Latin American Journal of Solids and Structures, 2013. 10. P. 197-222. 

\bibitem{hexa1}
I.V. Andrianov, V. V. Danishevskyy, A. Guillet and P. Pareige, Effective properties and micro-mechanical response of filamentary composite wires under longitudinal shear, European Journal of Mechanics A/Solids {\bf  24}, 195-206, 2005

\bibitem{square3}
I. V. Andrianov, G. A. Starushenko, V. V. Danishevskyy and S. Tokarzewski. Homogenization procedure and Pade approximants for effective heat conductivity of composite materials with cylindrical inclusions having square cross-section, Proceeding of Royal Society of London A {\bf 455}, 3401-3413, 1999
 
\bibitem{pade1}  G.A. Baker and P. Graves-Moris, Pad\'e Approximants (Cambridge
University, Cambridge, 1996)

\bibitem{bender} C.M. Bender and S.Boettcher, Determination of $f(\infty)$ from the asymptotoc series for $f(x)$ about $x=0$ J. Math.Phys. {\bf 35}, 1914-1921, 1994

\bibitem{moment}
D. J. Bergman, Dielectric-constant of a composite-material-problem in classical physics Phys. Rep.
{\bf 43},378-407, 1978

\bibitem{LB}
L. Berlyand and A. Kolpakov, Network approximation in the limit of small interparticle
distance of the effective properties of a high contrast random dispersed
composite, Arch. Ration. Mech. Anal. {\bf 159}, 179-227, 2001.

\bibitem{BN} 

L. Berlyand, A. Novikov, Error of the network approximation for densly packed composites with irregular geometry,
SIAM J. Math. Anal. {\bf 34}, 385-408, 2002

\bibitem{method}
R. Czapla, W. Nawalaniec and V. Mityushev, {\it Effective conductivity of random two-dimensional composites with circular non-overlapping inclusions}, 
Comput. Mat. Sci.{\bf 63}, 118-126, 2012.


\bibitem{christ}
R M.Christensen Mechanics of composite materials, Dover Publications (2005)
p.p. 1-384.

\bibitem{m3}
R. Czapla, W.Nawalaniec and V.Mityushev, Effective conductivity of random two-dimensional composites with circular non-overlapping inclusions, Comput. Mat. Sci.{\bf 63}, 118-126. 2012

\bibitem{trunc}
V. Derkach, S. Hassi, H, de Snoo, Truncated moment problems in the class of generalized
Nevanlinna functions, Math. Nachr.{\bf  285}, No. 14-15, 1741 – 1769, 2012 / DOI 10.1002/mana.201100268

\bibitem{rg1}
V. Dunjko, M. Olshanii,  A Hermite-Pade perspective on the renormalization group, with an application to the correlation function of Lieb-Liniger gas, J. Phys. A: Math. Theor. 44 055206 doi:10.1088/1751-8113/44/5/055206, 2011

\bibitem{puller} 
S. Gluzman, D.A. Karpeev,L.V.  Berlyand, Effective viscosity of
puller-like microswimmers: a renormalization approach. J. R. Soc. Interface 10: 20130720. http://dx.doi.org/10.1098/rsif.2013.0720, 2013

\bibitem{0} 
S. Gluzman, V. Mityushev, Series, Index and Threshold for Random 2D Composite, Arch. Mech., 67, 1, 75-93, Warszawa 2015

\bibitem{our1} 
S. Gluzman, V. Mityushev, W. Nawalaniec,
Cross-properties of the effective conductivity of the regular array of ideal conductors, Arch. Mech. {\bf 66}, 4, 287-301, 2014

\bibitem{padeour} 
S. Gluzman and V.I. Yukalov, Extrapolation of perturbation-theory expansions by self-similar approximants, European Journal of Applied Mathematics 25, 595 - 628 (2014).

\bibitem{cross}
S. Gluzman, V. Yukalov, Unified Approach to Crossover Phenomena, Physical Review E {\bf 58}, 4197-4209, 1998

\bibitem{corr} 
S. Gluzman, V.I.Yukalov, Self-similar extrapolation from weak to strong coupling, J.Math.Chem. {\bf 48}, 883-913, 2010

\bibitem{32}
S. Gluzman, V. Yukalov. Algebraic Self-Similar Renormalization in Theory
of Critical Phenomena, Physical Review E {\bf 55}, 3983-3999, 1997

\bibitem{fac1}  
S. Gluzman, V. I. Yukalov and  D. Sornette, {\it Self-Similar Factor Approximants}, 
Phys. Rev. E {\bf 67 (2)}, art. 026109, 2003.

\bibitem{4}
S. Gluzman, V.I. Yukalov, Self-similar continued root approximants, Physics Letters A {\bf 377}, 124-128, 2012.

\bibitem{moment1}
K. Golden, G.Papanicolaou, Bounds for effective parameters of heterogeneous media by analytic
continuation Commun. Math. Phys.{\bf 90}, 473-91 1983

\bibitem{gonch}
 A. A. Gonchar, Rational Approximation of Analytic Functions, Proceedings of the Steklov Institute of Mathematics,  {\bf 272}, Suppl. 2, pp. S44-S57, 2011
 
\bibitem{mom2}
P. Gonzales-Vera, O. Njastad,  Szego functions and multipoint Pade approximation, Journal of Computational and Applied Mathematics {\bf 32}, 107-116,  1990
 
\bibitem{hardy}
G.H. Hardy, M.Riesz, The general theory of Dirichlet's series,
Cambridge University Press (1915).


\bibitem{mom1}
E. Hendriksen, Moment methods in Two Point Pade Approximation, Journal of Approximation Theory {\bf 40}, 313-326, 1984
 
\bibitem{square1}
A. L. Kalamkarov,  I. V. Andrianov,  G. A. Starushenko, Three-phase model for a composite material with cylindrical circular inclusions. Part I: Application of the boundary shape perturbation method , International Journal of Engineering Science {\bf 78}, 154-177, 2014


\bibitem{square2} 
A. L. Kalamkarov,  I. V. Andrianov,  G. A. Starushenko, Three-phase model for a composite material with cylindrical circular inclusions, Part II: Application of Pad\'{e} approximants, International Journal of Engineering Science {\bf 78}, 178-191, 2014
 
\bibitem{kell1}
 J.B. Keller, Conductivity of a Medium Containing a Dense Array of Perfectly Conducting Spheres or Cylinders or Nonconducting Cylinders, Journal of Applied Physics {\bf 34}, 991-993, 1963


\bibitem{kell0} 
J.B.Keller, A Theorem on the Conductivity of a Composite Medium  J. Math. Phys.{\bf 5}, 548-549, 1964; doi: 10.1063/1.1704146


\bibitem{kr} 
M.G. Krein,  A.A. Nudel'man, The Markov moment problem and extremal problems
(in Russian), Nauka, Moscow, 1973; Translations of Mathematical Monographs
American Mathematical Society 50 (1977), Providence, RI.

\bibitem{complex}
A.I.  Markushevich, Theory of functions of a complex variable,1 , Chelsea (1977) (Translated from Russian) 
%MR0444912 Zbl 0357.30002

\bibitem{0}
J.C. Maxwell. 1873 Electricity and magnetism, p. 365, 1st . Oxford: Clarendon Press.

\bibitem{MCass} 
R. C. McPhedran, Transport Properties of Cylinder Pairs and of the Square Array of Cylinders, Proc.R. Soc. Lond. A {\bf 408}, 31-43, 1986, doi: 10.1098/rspa.1986.0108.

\bibitem{MCass1}
R.C. McPhedran, L, Poladian,G.W. Milton, Asymptotic studies of closely spaced, highly conducting cylinders. Proc. R. Soc. A {\bf 415}, 185-196, 1988.


\bibitem{m1} 
V. Mityushev, Steady heat conduction of a material with an array of cylindrical holes in the  nonlinear case, IMA Journal of Applied Mathematics {\bf 61}, 91-102, 1998.
 
\bibitem{m2} 
V. Mityushev, Exact solution of the $\mathbb R$-linear problem for a disk in a class of doubly periodic functions, J. Appl. Functional Analysis {\bf 2}, 115-127, 2007.

\bibitem{Mit2013b}
V. Mityushev, N. Rylko, {\it Maxwell's approach to effective conductivity and its limitations}, 
The Quarterly Journal of Mechanics and Applied Mathematics (2013); doi: 10.1093/qjmam/hbt003.

\bibitem{McPh} 
W.T. Perrins, D.R. McKenzie and R.C. McPhedran, Transport properties of regular array of cylinders. Proc.R. Soc.A {\bf 369}, 207-225 1979.


\bibitem{1} 

Rayleigh, On the influence of obstacles arranged in rectangular
order upon the properties of medium, Phil. Mag. {\bf 34}, 481-502, 1892.


\bibitem{Suetin}

S.P. Suetin, Pad\'{e} approximants and efficient analytic continuation of a power series, Russian Mathematical Surveys {\bf 57}, 43-141, 2002.

\bibitem{moment2}
 S. Tokarzewski, I. Andrianov, V. Danishevsky, G. Starushenko. Analytical continuation of asymptotic expansion of effective transport coefficients by Pade approximants, Nonlinear Analysis, {\bf 47}, 2283-2292, 2001
 
\bibitem{torkbook}
S. Torquato, {\it Random Heterogeneous Materials: Microstructure and Macroscopic Properties}, Springer-Verlag. New York, 2002.

\bibitem{torq} 
S.Torquato, F. H. Stillinger, Jammed hard-particle packings: From Kepler to Bernal and beyond Reviews of Modern Physica, {\bf 82}, 2634-2672, 2010. 

\bibitem{st1} 
G. Valent,  W. Van Assche, The impact of Stieltjes' work on continued fractions and
orthogonal polynomials: additional material, Journal of Computational and Applied Mathematics {\bf  65}, 419-447, 1995.

\bibitem{st} 
W. Van Assche, The impact of Stieltjes work on continued fractions and orthogonal polynomials, Thomas Jan Stieltjes Oeuvres Compl\`etes - Collected Papers", G. van Dijk, ed., Springer, 1993, pp. 5-37,
Report number:	OP-SF 9 Jul 1993.

\bibitem{Wall}
H.S. Wall, Analytic Theory of Continued Fractions, Chelsea Publishing Company, Bronx N.Y. 1948

\bibitem{fac2} V. I. Yukalov, S. Gluzman, D. Sornette, {\it Summation of Power Series by Self-Similar Factor Approximants}, 
Physica A, {\bf 328}, 409-438, 2003. 

\bibitem{cross1}

V. Yukalov, S. Gluzman, Self Similar Crossover in Statistical Physics, Physica A {\bf 273}, 401-415, 1999

\bibitem{superexp}
V. Yukalov, S. Gluzman, Self-Similar Exponential Approximants, Physical Review E. {\bf 58}, 1359-1382,1998

%\bibitem{hardy}
%G.H. Hardy, M.Riesz, The general theory of Dirichlet's series,
%Cambridge University Press (1915).

%\bibitem{complex}
%A.I.  Markushevich, Theory of functions of a complex variable,1 , Chelsea (1977) (Translated from Russian) 
%%MR0444912 Zbl 0357.30002




\end{thebibliography}
\end{document}